\documentclass[trackchanges,twocolumn]{aastex7}

\usepackage{amssymb}
\usepackage{amsmath}

             \font\sevenrm=cmr7

\newcommand\red[1]{{\color{black}#1}}
\def\fsc{\alpha_{\hbox{\sevenrm f}}} 

\def\rmax{r_{\hbox{\sevenrm max}}}
\def\RNS{R_{\hbox{\sevenrm NS}}}

\def\omegaB{\omega_{\hbox{\sevenrm B}}}
\def\sigmaeff{\sigma_{\hbox{\sevenrm eff}}}
\def\neff{n_{\hbox{\sevenrm eff}}}
\def\Nptoq{N_{p\rightarrow q}}

\newcommand{\xGamma}{x_{\Gamma}}
\def\nmax{n_{\hbox{\sevenrm max}}}
\def\thetacool{\theta_{\hbox{\sevenrm cool}}}
\def\thetalock{\theta_{\hbox{\sevenrm lock}}}
\def\gammalock{\gamma_{\hbox{\sevenrm lock}}}
\def\gammaSC{\gamma_{\hbox{\sevenrm SC}}}
\def\rres{r_{\hbox{\sevenrm res}}}

\def\Eparallel{\boldsymbol{E}_\parallel}

\def\phires{\phi_{\hbox{\sevenrm res}}}
\def\rpl{r{\hbox{\sevenrm pl}}}

\definecolor{ao(english)}{rgb}{0.0, 0.5, 0.0}

\begin{document}

\title{Resonant Inverse Compton Scattering and Hard X-ray Emission in Magnetar Magnetospheres}

\author[orcid=0000-0002-9705-7948,sname='']{Kun Hu}
\affiliation{Washington University in St. Louis, Department of Physics and McDonnell Center for the Space Sciences, St. Louis, MO 63130, United States}
\email[show]{hkun@wustl.edu} 

\author[orcid=0000-0002-4402-1343,sname='']{Nicholas Rackers}
\affiliation{Washington University in St. Louis, Department of Physics and McDonnell Center for the Space Sciences, St. Louis, MO 63130, United States}
\email[]{n.rackers@wustl.edu} 

\author[orcid=0000-0002-4738-1168,sname='']{Alexander Y. Chen}
\affiliation{Washington University in St. Louis, Department of Physics and McDonnell Center for the Space Sciences, St. Louis, MO 63130, United States}
\email[]{cyuran@wustl.edu} 

\begin{abstract}

Magnetars are a subclass of neutron stars with ultra-strong surface magnetic fields. Some magnetars exhibit persistent hard X-ray emission, characterized by power-law tails with photon indices around 1--1.5, extending from ${\sim}$10 keV to several hundred keV. The leading explanation for this hard X-ray component is resonant Compton scattering, in which the thermal seed photons are upscattered by relativistic electron-positron pairs flowing along magnetic field lines in the magnetosphere.  
In this work, we adopt the pair outflow framework of the magnetar magnetosphere and calculate the resonant Compton scattering opacity, as well as the spectrum and polarization of the upscattered emission. 
We find that resonant cooling can substantially modify the magnetospheric plasma density and impose strong optical depth constraints on the hard X-ray emission regions. Under the viewing geometry inferred from \emph{IXPE}, an equatorial twist near the stellar surface provides a viable configuration for the \emph{NuSTAR} hard X-ray spectrum of 4U 0142+61, while a polar-twist geometry is disfavored.
Joint spectral, timing, and polarimetric modeling will be essential for distinguishing between the magnetospheric scattering geometries and understanding the physical properties of the pair plasma.

\end{abstract}



\section{Introduction} 

Magnetars are isolated neutron stars characterized by
{their bright X-ray emission}
that often exceeds their spin-down power. Unlike ordinary rotation-powered pulsars, magnetar emission is believed to be powered primarily by the dissipation and decay of their ultra-strong magnetic fields \citep[][]{1992ApJ...392L...9D,1995MNRAS.275..255T}. Measurements of the periods and period derivatives indicate that these objects host exceptionally strong surface dipole fields, $B_p\gtrsim 10^{14}\,\mathrm{G}$, exceeding the quantum critical field strength $B_{cr}=m_e^2c^3/(e\hbar)\approx 4.4\times10^{13}\,\mathrm{G}$. Such extreme fields support the picture that the release and dissipation of magnetic energy powers their high-energy emission. 

Soft X-ray emission (0.5–10 keV) from magnetars is typically modeled as a thermal blackbody component plus a steep power law with photon index $\Gamma\sim2\text{--}4$ \citep[][]{2008A&ARv..15..225M}. The thermal component is generally associated with radiation from the neutron-star surface, while the power-law tail is commonly attributed to surface photons being reprocessed via repeated resonant cyclotron Compton scattering by charged particles in a twisted, current-carrying magnetosphere \citep[][]{2006MNRAS.368..690L,2008MNRAS.386.1527N}. In some sources, two blackbodies provide a better description, consistent with emission from multiple surface temperature regions \citep[][]{2015ApJ...808...32T}. 

Some magnetars also display a persistent hard X-ray power-law tail above around 10 keV, extending to around 100--200 keV \citep[][]{2006A&A...451..587D,2006A&A...449L..31G,2006ApJ...645..556K}. In contrast to the soft-band tail, this hard component is spectrally flat, with typical photon indices $\Gamma\sim 1\text{--}2$, and is expected to turn over above a few hundred keV as required by the upper limits from \emph{COMPTEL} and \emph{Fermi}--LAT at higher energies \citep[][]{2010ApJS..187..460A,2017ApJ...835...30L}. Long-term monitoring indicates that the hard X-ray tail is generally stable over multi-decade timescales, although its flux may evolve following bursting activities in some sources \citep[][]{2014RAA....14..673W,2025AN....34640109P}.

The leading theoretical interpretation of the persistent hard X-ray tail in magnetars invokes resonant Compton scattering (RCS) in a twisted, current-carrying magnetosphere \citep[][]{2007Ap&SS.308..109B,2007ApJ...660..615F,2013ApJ...762...13B,2018ApJ...854...98W}. In this framework, crustal motion shears the anchored magnetic footpoints, imparting a global or localized twist to the external magnetosphere and storing free energy, which subsequently dissipates over time. The twisted configuration is maintained by field-aligned electric currents, with a characteristic current density $\boldsymbol{j}\propto \nabla\times\boldsymbol{B}$. To conduct the current, plasma is extracted from the surface and/or produced in the magnetosphere, and is accelerated along magnetic field lines. These charged particles then resonantly upscatter the thermal X-ray photons at the cyclotron frequency (in their rest frame) into hard X-rays. The energy of the outgoing photons, compared with the seed photons in the lab frame, is boosted by a factor of $\gamma^2$, where $\gamma$ is the Lorentz factor of the scattering electron. This model has been compared directly with \emph{Nuclear Spectroscopic Telescope Array (NuSTAR)} observations of several persistent magnetars and successfully constrained the magnetic and viewing geometry of these magnetars~\citep[][]{2014ApJ...786L...1H,2015ApJ...807...93A}.

Alternative pictures have been proposed to explain the hard X-ray emission. Recent pair cascade simulations suggest that the magnetic pair creation and photon splitting strongly reprocess the primary resonantly scattered photons, and the synchrotron emission from the secondary pairs could dominate the observed hard X-ray emission \citep[][]{2025ApJ...991..178H,2025ApJ...985L..35S}. A second class is the annihilation bremsstrahlung picture, in which the annihilation of a pair plasma in supercritical fields produces a bremsstrahlung-like spectrum covering the hard X-ray band \citep[][]{2020ApJ...904..184T,2025ApJ...986..173Z}.

Recent \emph{Imaging X-ray Polarimetry Explorer} (\emph{IXPE}) observations of magnetars have revealed strong, energy-dependent linear polarization in the 2–8 keV band \citep[][]{2022Sci...378..646T,2023ApJ...944L..27Z,2023ApJ...954...88T,2024MNRAS.52712219H,2024Galax..12....6T,2025ApJ...985L..34R,2026ApJ..1002..102T}. The observed high polarization degree (PD) from some sources strongly indicates the presence of QED vacuum birefringence, in which virtual pairs make the magnetized vacuum behave like a birefringent medium and the polarization state of photons freezes out within a polarization limiting radius \citep{2003MNRAS.342..134H,2003ApJ...588..962L}. Although \emph{IXPE} probes the soft X-ray band, the measured polarization angle (PA) variation, interpreted with the rotating vector model, provides an independent constraint on the star’s viewing and magnetic geometry. This, in turn, helps anchor models for where and how the persistent hard X-ray tail is produced.

In this paper, we revisit the resonant scattering model with pair outflow for magnetar hard X-ray emission and attempt to use the magnetic inclination and viewing angles inferred from \emph{IXPE} observations under the rotating vector model to constrain the possible field line twist geometry. 
\red{While RCS radiative transfer is a well-established technique, previous models often decouple the magnetospheric plasma parameters from the scattering process. The central novelty of our work is the physically self-consistent integration of the pair-outflow prescription, continuous pair plasma deceleration via resonant scattering, and the resulting hard X-ray production.}
In Section \ref{sec:resonant_scattering} we briefly summarize the basics of RCS in the Thomson regime and the scattering geometry in the magnetosphere. In Section \ref{sec:pair_outflow}, we compute the electron cooling due to RCS, and the resulting magnetospheric opacities adjusted for the plasma flow. {In particular, we demonstrate that} {the pair density required to explain the hard X-ray luminosity could render much of the magnetosphere opaque to surface photons}. {In Section~\ref{sec:spectrum}, we specialize to some specific twist configurations and compute the resulting X-ray spectra from RCS, and present the corresponding polarization and pulse profiles.}  
{Finally, Section~\ref{sec:conclusion} summarizes our main findings and discusses potential issues and future prospects.}

\section{Resonant Compton scattering in magnetar magnetospheres} 
\label{sec:resonant_scattering}

{We start by summarizing the formalism and main ingredients necessary for computing the resonant Compton scattering process in the magnetospheres of magnetars.} In this work, we treat the resonant scattering in the Thomson region, neglecting the Klein–Nishina reduction and electron recoil. This approximation is appropriate when the local magnetic field is predominantly subcritical and the scattering electrons are mildly relativistic, which is generally expected in magnetar magnetospheres away from the immediate stellar surface, where hard X-ray is produced. 
Treatments incorporating relativistic kinematics and QED scattering cross sections can be found in \cite{2018ApJ...854...98W}.

\subsection{Magnetic Thomson Cross Section} \label{sec:cross_sec}

In the magnetospheres of magnetars, vacuum polarization typically dominates plasma dispersion, and the photon polarization is well described in terms of two normal modes, $\perp$ and $\parallel$ \citep[e.g.,][]{2006RPPh...69.2631H}.
These modes are approximately orthogonal and linearly polarized, defined with respect to the plane formed by the photon wavevector $\boldsymbol{k}$ and the local magnetic field $\boldsymbol{B}$. 
The $\perp$ mode has its electric field $\boldsymbol{E}$ perpendicular to the $\boldsymbol{k}$-$\boldsymbol{B}$ plane, while the $\parallel$ mode has its $\boldsymbol{E}$ vector within it.
The scattering cross section exhibits a resonance when the photon frequency $\omega$ matches the local gyro-frequency $\omegaB =eB/(m_e c)$. Near the resonance, the azimuthal $\phi$ dependence of the cross section vanishes, allowing the cross sections to be written in the simplified form:
\begin{equation} \label{eq:cross_sec_reso}
\begin{split}
\frac{d\sigma_{\parallel\rightarrow\parallel}}{d\Omega_f}  \,=\, & \frac{r_0^2}{4}\frac{\omegaB^2}{(\omega-\omegaB)^2+\Gamma^2/4}\cos^2{\theta_i}\cos^2{\theta_f},\\
\frac{d\sigma_{\parallel\rightarrow\perp}}{d\Omega_f}  \,=\, & \frac{r_0^2}{4}\frac{\omegaB^2}{(\omega-\omegaB)^2+\Gamma^2/4}\cos^2{\theta_i},\\
\frac{d\sigma_{\perp\rightarrow\parallel}}{d\Omega_f}  \,=\, & \frac{r_0^2}{4}\frac{\omegaB^2}{(\omega-\omegaB)^2+\Gamma^2/4}\cos^2{\theta_f},\\
\frac{d\sigma_{\perp\rightarrow\perp}}{d\Omega_f}  \,=\, & \frac{r_0^2}{4}\frac{\omegaB^2}{(\omega-\omegaB)^2+\Gamma^2/4}.
\end{split}
\end{equation}
Here $r_0 = e^2/(m_e c^2)$ is the classical electron radius, $\Gamma $ 
is the line width due to the finite decay time, $\theta_i$ and $\theta_f$ are the incoming and outgoing angles of the photons with respect to the field direction in the electron rest frame (ERF).
We can rewrite the cross sections in a compact matrix format:
\begin{equation}
    \overleftrightarrow{{\frac{d\sigma}{d \Omega} }}= 
    \begin{bmatrix}
   \displaystyle\frac{d\sigma_{\parallel\rightarrow\parallel}}{d \Omega} &  \displaystyle\frac{d\sigma_{\perp\rightarrow\parallel}}{d \Omega} \\
   \displaystyle\frac{d\sigma_{\parallel\rightarrow\perp}}{d \Omega} & \displaystyle \frac{d\sigma_{\perp\rightarrow\perp}}{d \Omega} \\
\end{bmatrix} = 
  \frac{r_0^2\, b^2}{4\,x_{\Gamma}}
    \frac{x_{\Gamma}}{(x-b)^2+x_{\Gamma}^2/4}
    {\cal P}_{p\rightarrow q},
\end{equation}
where $x = \hbar \omega/(m_ec^2)$, $b = \hbar \omega_B/(m_ec^2)=B/B_{cr}$, and $x_{\Gamma} = \hbar \Gamma/(m_ec^2) $.
Here 
\begin{equation}
    {\cal P}_{p\rightarrow q}=
    \begin{bmatrix}
   \cos^2{\theta_i} \cos^2{\theta_f} &   \cos^2{\theta_f} \\
    \cos^2{\theta_i} &  1 \\
\end{bmatrix},
\label{eqn:M_matrix}
\end{equation}
and $p,\, q =\, \parallel$ or $\perp$, denoting the polarization states of incident and outgoing photons.
When $x_\Gamma \ll 1$, the Lorentzian function inside the cross sections can be approximated by a $\delta$ function
\begin{equation}
    \overleftrightarrow{{\frac{d\sigma}{d \Omega} }}
    = \sigmaeff\delta(x-b) {\cal P}_{p\rightarrow q},
\end{equation}
and we define
\begin{equation}
    \sigmaeff = \frac{\pi b^2}{2 x_\Gamma}  r_0^2 .
\end{equation}

The dimensionless line width $\xGamma$ can be expressed as a function of the field strength. A useful analytic approximation is provided in Equation (A3) of \cite{2016MNRAS.461..877V}:
\begin{equation}
    \xGamma 
    \approx \fsc \frac{(2/3)b^2 (1-1/e)}{[(2b^2/3)^{1/q}+[1-1/e]^{1/q}]^q}, \quad q = 2.105,
    \label{eqn:x_Gamma}
\end{equation}
\red{where we adopt $q=2.105$, which gives slightly better agreement with the exact expression than the value $q=2\pi/3$ presented by \citet{2016MNRAS.461..877V}. This approximation} accurately captures the asymptotic behavior of $\xGamma$ in both strong $(\xGamma\approx \fsc(1-1/e)\approx0.0046)$ and weak fields $(\xGamma\approx (2/3)\fsc b^2)$, with $\fsc = e^2/(\hbar c)$ being the fine structure constant. 
Note this low field value accounts for the spin-averaged cyclotron line width, which is exactly half of the classical non-relativistic value of $(4/3) \fsc b^2$ \citep[e.g., see][]{2005ApJ...630..430B}. 
In this work, we perform calculations using both the $\xGamma$ prescription in Equation~(\ref{eqn:x_Gamma}) and the weak-field value $(2/3)\fsc b^2$. Since the resulting spectra show only minor differences, we will neglect the field strength dependence of the line width and adopt $(2/3)\fsc b^2$ unless stated otherwise. 

\subsection{Resonant Compton Scattering Rate}

Building on \cite{1990ApJ...360..197D} and \cite{1989ApJ...343..277H}, the photon scattering rate for a single electron in the observer frame (OF) can  be written as
\begin{equation}
\begin{split}
    \frac{d\Nptoq}{dt\, dx_f\,  %
    d\Omega_f dN_e}
    = \frac{c}{\gamma (1-\beta \cos{\theta_f})} \times &\\
    \int dx_i \int 
    d\Omega_i
    \,\frac{dn_s}{dx_i d \Omega_i}\frac{d \sigma'}{dx'\, 
    d\Omega'_f
    }& (1-\beta \cos{\theta_i}),
    \end{split}
    \label{eqn:dNph_dtdxdOdNe}
\end{equation}
where $\beta$ is the dimensionless velocity of the electron, $\gamma$ is the Lorentz factor, and $\theta_{i}$ and $\theta_{f}$ are the incident and outgoing angles, respectively. The integration is over $x_i$, the normalized incoming photon energy in the OF, and $\Omega_i$ which is the solid angle subtended by the emitter.  
Here $p$ and $q$ label the incident and scattered photon polarization modes, i.e., $\perp$ or $\parallel$. The $(1-\beta\cos{\theta_f})$ factor arises when transforming the photon scattering rate from the electron rest frame (ERF) to the OF, and the $(1-\beta\cos{\theta_i})$ factor {accounts for the transformation of the incoming photon from the OF to the ERF.}
Primes denote quantities evaluated in the ERF. 
Note the $\perp$ and $\parallel$ mode polarizations do not mix under Lorentz boost along the $\boldsymbol{B}$ direction. The $\perp$ mode polarization vector does not change under Lorentz boost, while the $\parallel$ mode polarization vector remains inside the $\boldsymbol{k}$-$\boldsymbol{B}$ plane after a gauge transformation.

Equation (\ref{eqn:dNph_dtdxdOdNe}) is essentially a linear expression, in which the total scattering rate is assumed to be the single-electron rate multiplied by the number of available electrons. This treatment is valid when the medium is optically thin to scattering \red{(see Appendices~{\ref{app:scattering}} and {\ref{app:optical_thin}})}. 
If the electron density is high enough that most of the photons are likely to scatter, but the photon flux is insufficient to illuminate all electrons, the scattering rate is no longer proportional to the total number of electrons. In that case, a proper radiative transfer calculation is required \citep[see, e.g.][]{2006MNRAS.368..690L}. 

In the Thomson regime, 
the differential cross section reduces to 
\begin{equation}
    \frac{d \sigma'}{dx'\, 
    d\Omega'} = 
     \left(\frac{d \sigma'}{ 
    d\Omega'}\right) \delta(x'_i-x'_f)
    \approx\sigmaeff {\cal P}_{p \rightarrow q}  \delta(x'_i-x'_f) \delta(x'-b).
    \label{eqn:cross_section}
\end{equation}
Here we 
omit 
the double-sided arrow notation for simplicity, {noting that the matrix $\mathcal{P}_{p \rightarrow q}$ explicitly shows up in this cross-section}. 
In Equation~(\ref{eqn:cross_section}), the first $\delta$ function enforces energy conservation in the non-relativistic regime, so the scattered photon energy is unchanged in the ERF. We use it to simplify the integral over the incident photon energy. 
The second $\delta$ function 
approximates the resonant cross section as a narrow line at the cyclotron frequency.
We will use it to simplify the volume integral over the scattering region. {Since the line width $x_\Gamma$ is small $(\xGamma\lesssim0.0046)$ throughout the magnetosphere, this is a good approximation, especially for hard X-ray emission.}

\subsection{Volume-Integrated RCS Spectrum}
\label{sec:RCS_in_field_loops}


The observed spectrum can be obtained by integrating the specific photon scattering rate in Equation~(\ref{eqn:dNph_dtdxdOdNe}) over the scattering volume of the magnetosphere, i.e,
\begin{equation}
    \frac{d\Nptoq}{dt\, dx_f\, 
    d\Omega_f} = \int n_e(r,\theta,\phi) d V \left(\frac{d\Nptoq}{dt\, dx_f\, 
    d\Omega_f\,d N_e } \right).
    \label{eqn:spec_vol_inte}
\end{equation}
In general, Equation~(\ref{eqn:spec_vol_inte}) requires the summation of the polarization vectors, since their polarization states are determined locally at the scattering site{, which may lead to cancellations when the photons reach the observer}. In the surroundings of magnetars, however, photon polarization vectors typically evolve adiabatically with the local field direction during the photon propagation until it reaches the ``polarization limiting radius" $\rpl$ \citep{2003MNRAS.342..134H,2003ApJ...588..962L}. {The $\perp$ and $\parallel$ modes do not mix, and the final polarization vector direction is mostly determined by the magnetic field direction at $\rpl$~\citep{2011ApJ...730..131F}.} For magnetar field strength, $\rpl$ generally exceeds a few hundred stellar radii, where the photon trajectories from different scattering spots are almost radial and experience the same field structures. 
Therefore, we directly work with photon polarization modes $\perp$ or $\parallel$ and sum the number of photons in each mode to produce the polarized emission at infinity.

{In order to evaluate the integral in Equation~\eqref{eqn:spec_vol_inte}, we specialize to a coordinate system $(r_\mathrm{max}, \theta, \phi)$ where $r_\mathrm{max}$ marks the maximum extent of a given magnetic field line. Although this is not an orthogonal coordinate system (its metric is not diagonal), it is useful since it offers a particularly convenient way to integrate over selected field line bundles.} We first integrate the scattering rate over the azimuthal angle $\phi$ with respect to the magnetic axis to exploit the $\delta(x'-b)$ function in Equation~(\ref{eqn:cross_section}). Therefore, 
\begin{equation}
\begin{split}
    \frac{dN_{p \rightarrow q}}{dt\, dx_f\,   
    d\Omega_f d\rmax d\theta}&(x_f,\rmax,\theta,\Omega_f) = \int d\phi\, n_e \rmax^2\sin^7{\theta} \\
    \times \RNS^3\frac{c}{\gamma^2 (1-\beta \mu_f)}& \int d\Omega_i
    \frac{dn}{dx_i d \Omega_i}\,
    \sigmaeff{\cal P}_{p\rightarrow q}\sum_{\phi_{res} }\frac{\delta(g(\phi))}{|g'(\phi)|}\\
    = \sum_{\phires}
    n_e \rmax^2\sin^7{\theta} & \RNS^3 \frac{c \sigmaeff}{\gamma^2 (1-\beta \cos{\theta_f})}  \frac{1}{|g'(\phi)|}\\
    \times & 
   \int d\Omega_i \frac{dn}{dx_i d \Omega_i} {\cal P}_{p\rightarrow q}. 
\end{split}
\label{eqn:dNph_dt_dxf_dOf_drmax_dtheta}
\end{equation}
Here we adopt the Jacobian of the volume integration $dV=\rmax^2\sin^7{\theta}\RNS^3 d\rmax d\theta d\phi$, and define
\begin{equation}
    g(\phi)=x_f\gamma(1-\beta\mu_f)-b.
    \label{eqn:g_phi}
\end{equation}
Alternatively, we can use the field loop footpoint colatitude $\theta_{\rm foot}$ instead of $\rmax$, and the volume element $dV = r^3\sin{\theta_{\rm foot}}\cos{\theta_{\rm foot}}\RNS^3 drd\theta_{\rm foot} d \phi/\cos{\theta}$. 
The outgoing scattering angle $\mu_f$ can be related to the scattering locale $(\rmax,\theta,\phi)$ and the viewing colatitude with respect to the magnetic axis, $\theta_v$, via 
\begin{equation}
    \mu_f = \frac{3\sin{\theta}\cos{\theta}\sin{\theta_v}{\cos{\phi}}+(3\cos^2{\theta}-1)\cos{\theta_v}}{\text{sgn}(\pi/2-\theta)\,\sqrt{3\cos{\theta}^2+1}}. 
    \label{eqn:muf_rm_theta_phi}
\end{equation}
From Equation~(\ref{eqn:muf_rm_theta_phi}), one can quickly obtain
\begin{equation}
    g'(\phi)=x_f\gamma\beta \frac{3\sin{\theta}\cos{\theta}\sin{\theta_v}}{\text{sgn}(\pi/2-\theta)\sqrt{3\cos{\theta}^2+1}} \sin{\phi}.
\end{equation}
The $\text{sgn}(\pi/2-\theta)$ function takes into account the assumption that electrons are moving from the stellar surface towards the equatorial area for both stellar hemispheres. 

Given the observed energy $x_f$ and the scattering site at $(\rmax,\theta)$, the resonance condition $g(\phi)=0$ restricts the outgoing photon to lie on a cone of constant $\mu_f$ around the local magnetic-field direction. The final propagation direction $\theta_v$ then selects the corresponding azimuth $\phires$ at which the resonance is satisfied, namely:

\begin{equation}
\begin{split}
    {\cos{\phires}}& = \frac{1}{3\sin{\theta}\cos{\theta}\sin{\theta_v}}\\
    \times &\bigg\{\sqrt{3\cos{\theta}^2+1}\left(\frac{1}{\beta}-\frac{b}{\beta\gamma x_f}\right)\,\text{sgn}(\pi/2-\theta) \\
   & \qquad \qquad \qquad-(3\cos^2{\theta}-1)\cos{\theta_v}\bigg\}.
\end{split}
\end{equation}
Given the final photon direction $\theta_v$ and the meridional coordinates $(\rmax,\theta)$ for the scattering, Equation~(\ref{eqn:dNph_dt_dxf_dOf_drmax_dtheta}) gives the scattered photon spectrum explicitly. 
The total spectrum from any azimuthally symmetric magnetospheric volume then follows by integrating Equation~(\ref{eqn:dNph_dt_dxf_dOf_drmax_dtheta}) over $\rmax$ (or $\theta_f$) and $\theta$.
It can then be extended to an arbitrary emitting volume by filtering the resonant $\phires$ in Equation~(\ref{eqn:dNph_dt_dxf_dOf_drmax_dtheta}).
One may also use the $\delta(x'-b)$ function to eliminate either $\rmax$ or $\theta$ in the volume integral. In practice, however, the resonant values of $\rmax$ or $\theta$ do not have closed-form expressions, requiring numerical root finding to solve the integral.

One caveat of Equation~(\ref{eqn:dNph_dt_dxf_dOf_drmax_dtheta}) is that it requires $\beta$ to be finite. When $\beta\rightarrow0$, $\mu_f$ becomes undetermined from Equation~(\ref{eqn:g_phi}).
In this limit, we instead use:

\begin{equation}
\begin{split}
    \frac{dN_{p\rightarrow q}}{dt\, dx_f\,   
    d\Omega_f d\rmax d\theta} (x_f,& r_m,\theta,\Omega_f) = 
    \int d\phi\, n_e \rmax^2\sin^7{\theta}\\
    \times\, \RNS^3 \frac{c}{\gamma^2(1-\beta\cos{\theta_f})}&\int d\Omega_i
    \frac{dn}{dx_i d \Omega_i}\,
    \sigmaeff{\cal P}_{p \rightarrow q}  \delta(x_f-b)\\
    \approx  n_e \rmax^2\sin^7{\theta}\, c \RNS^3 & \sigmaeff  \frac{1}{\gamma^2} \int d\Omega_i
    \frac{dn}{dx_i d \Omega_i}\,
    \\
    \times 
    \int & d\phi\,\frac{{\cal P}_{p\rightarrow q}}{(1-\beta\mu_f)} \frac{\int_{x_{f1}}^{x_{f2}}\delta(x_f-b)}{x_{f2}-x_{f1}}.
\label{eqn:dNph_dt_dxf_dOf_drmax_dtheta_2}
\end{split}
\end{equation}
In Equation~(\ref{eqn:dNph_dt_dxf_dOf_drmax_dtheta_2}), we directly integrate the polarization-dependent ${\cal P}_{p\rightarrow q}$ factor over $\phi$ using Equation~(\ref{eqn:M_matrix}) and (\ref{eqn:muf_rm_theta_phi}). We eliminate the $\delta(x'-b)$ factor by performing a localized integration inside the boundaries of the observed energy bin $(x_{f1},x_{f2})$. In practice, this can be done routinely using an array selection function with the condition $x_{f1}<b<x_{f2}$.

The last ingredient in Equation~(\ref{eqn:spec_vol_inte}) is the total scattering-charge density $n_e=n_++n_-$ along each meridional field line. It can be written in terms of the current density $j_e$ as
\begin{equation}
    n_e = \frac{{\cal M}j_e}{c\beta e} = 
     \frac{{\cal M}I}{c \beta e \Delta A},
\end{equation}
where $I$ is the electric current, $\Delta A$ is the flux-tube cross section area, {$\beta = v/c$ is the dimensionless velocity of the electron flow,} and ${\cal M}$ is the multiplicity {over the minimum density to conduct the current}. According to \cite{2013ApJ...777..114B}, self-regulating pair production increases ${\cal M}$ from 1 to $\sim 100$ near the stellar surface. In the sub-critical regime further out, pair creation is inefficient, so we assume pair production finishes close to the star and take ${\cal M}$ to be constant over most of the field loop. With this assumption, charge conservation implies that the pair-plasma density scales with the local field strength as
\begin{equation}
    n_e = n_0 \frac{B}{\beta B_{\rm foot}},
    \label{eqn:n_e}
\end{equation}
where $B_{\rm foot}$ is the field strength at the foot point of the field tube and $n_0$ is the number density normalization. In principle, the absolute density could be fixed equivalently by ${\cal M}j_e$ or by $n_0$. In this work, we adopt $n_0$ as a free normalization parameter for simplicity, and implicitly absorb any dependence on ${\cal M}j_e$ into $n_0$.

The value of $n_0$ can be estimated using the observed X-ray flux. Following \cite{2007Ap&SS.308..109B}, the number density can be expressed as $n_e\sim 3\times10^{17} L_{X,35} /(\epsilon_{\rm rad}\langle\gamma_e\rangle)$ cm$^{-3}$ where $L_{X,35}$ is the hard X-ray luminosity in units of $10^{35}$ erg/s, $\epsilon_{\rm rad}$ is the radiation efficiency and $\langle\gamma_e\rangle$ is the mean electron Lorentz factor. Substituting $L_{X,35}\sim2$ and $\epsilon_{rad}\langle\gamma_e\rangle\sim100$ keV$/m_ec^2\approx0.2$ yields $n_e\sim3\times 10^{18}$ cm$^{-3}$. This is a reasonable value, and we will show in the following sections that adopting $n_0\sim10^{18}$ cm$^{-3}$ indeed produces an integrated spectrum consistent with the observed flux. Finally, we impose an upper limit $\nmax\sim 5\times10^{22}$ cm$^{-3}$ on the electron density to avoid divergence as $\beta$ approaches 0.

\section{Pair Plasma Outflow in the Magnetosphere}
\label{sec:pair_outflow}
In this work we adopt the pair outflow model for magnetar hard X-ray emission. In this picture, an active bundle of closed magnetic field lines is twisted by surface motion, demanding a strong current to flow along the field lines. To conduct the current, the twisted flux tube sustains a strong discharge that generates abundant $e^\pm$ pairs and accelerates them near the stellar surface. The pairs then stream along the magnetic field, with flows from the two hemispheres propagating outward toward the magnetic equator. RCS of thermal X-ray photons imposes strong radiative drag, regulating the cooling of the pairs and converting a significant fraction of the outflow energy into the observed hard X-ray component.

{In an optically thin magnetosphere, the thermal X-ray photons come from the stellar surface, and serve as the seeds for resonant Compton scattering. However, as pointed out by~\citet{2013ApJ...777..114B}, this optically thin assumption may not always be correct when electron cooling is taken into account. In this section, we will compute the evolution of the electron bulk Lorentz factor along the magnetic field line due to RCS cooling under different assumptions of the background photon distribution. Then, we will use this result to compute the electron density and RCS optical depth in the twisted magnetosphere. The goal of this section is to explicitly point out the realms of applicability of various assumptions and approximations regarding electron flow and RCS in the magnetar magnetosphere.}

\subsection{Electron Cooling}
\label{sec:Electron_Cooling}

In this section, we estimate the radiative cooling of the electron-positron pair plasma in the dipolar magnetosphere by computing the photon momentum transfer. We follow \cite{2013ApJ...777..114B} and assume that the momentum component along the magnetic field direction is conserved across each scattering event. 
We assume the pair plasma is cold. 
In general, a velocity difference between positron and electron $\beta_+-\beta_-\sim2\beta/({\cal M}+1)$ is required to sustain the electric current.
Since ${\cal M}\sim100\gg1$, the velocity difference is much smaller than the bulk motion velocity. Therefore we assume $\beta_+\sim \beta_-\sim\beta$ and neglect the $\Eparallel$ that would be needed to maintain the velocity separation. In this case, electrons and positrons contribute equally to the RCS process.
In the ERF, the change in electron momentum along the field after one scattering is
\begin{equation}
    \Delta p' = m_e c (x_i'\mu_i'-x_f'\mu_f').
\end{equation}
The time derivative of the momentum is the same in the OF and the ERF because the energy change $dE' =m_ec^2(x_i'-x_f')$ in the ERF is negligible {in the Thomson regime}. Therefore, 
\begin{equation}
    \frac{dp}{dt}=\frac{\gamma(dp'+\beta dE')}{\gamma dt'}\approx\frac{dp'}{dt'}.
\end{equation}
{This can be transformed into an equation for the time derivative of the electron Lorentz factor $\gamma$:}
\begin{equation}
     \frac{dp}{dt}=m_ec\frac{d(\gamma\beta)}{dt}=\frac{1}{\beta}\frac{d\gamma}{dt}m_ec .
\end{equation}
Finally, the time derivative of the electron 
{Lorentz factor}
due to a continuous photon shower can be expressed as
\begin{equation}
     \begin{split}
        &\frac{d\gamma_p}{dt} = \frac{\beta}{m_ec}\frac{dp'_p}{dt'} 
    \\
    &= \beta c 
    \int dx_f'  d \Omega'_f dx_i' d \Omega_i'\, (x_i'\mu_i'-x_f'\mu_f')\frac{dn'}{dx_i'd \Omega_i'}\frac{d \sigma'}{dx' 
    d\Omega'_f} \\
    & = \beta c  \sigmaeff  b \int d\Omega_i\mu_i' 
    \left.\frac{dn}{dx_i d \Omega_i}\right|_{x_i = x_r}
    \int d\Omega_f'%
    \sum_{q=\perp,\parallel} {\cal P}_{p\rightarrow q}
    \\
    &=\frac{16\pi}{3}\beta c  \sigmaeff   b \int d\Omega_i \left.\frac{dn}{dx_i d \Omega_i}\right|_{x_i = x_r}
    \begin{bmatrix}
         \mu_i' \quad \text{for}\quad p = \perp\\
        \mu_i'^3\; \text{for}\quad p = \parallel
    \end{bmatrix}.
    \\
    \end{split}
    \label{eqn:dgamma_dt}
\end{equation}
Here, the $x'_f\mu'_f$ term in the integrand vanishes after the integration due to the symmetry of the differential cross section in the ERF. The energy $x_r=b/[\gamma(1-\mu_i\beta)]$ is the resonant energy measured in the OF for the seed photons. 
In Equation~(\ref{eqn:dgamma_dt}), $\mu'_i=(\mu_i-\beta)/(1-\mu_i\beta)$ is the cosine of the incident angle measured in the ERF, and it determines the sign of the time derivative of $\gamma$. %

For electrons streaming along magnetic field lines, both the scattering location $r$ and the path length traveled by an electron $s$ can be parameterized by the magnetic colatitude $\theta$, measured with respect to the magnetic axis, and $\rmax$, the maximum radius of a given field line, i.e.,
\begin{equation}
    r = \rmax \sin^2{\theta},\quad
    ds = \rmax \sqrt{1+3\cos^2{\theta}} \sin{\theta} d\theta,
\end{equation}
where $r$, $\rmax$ and $s$ are unitless variables rescaled by the stellar radius $\RNS$. 
If the electron motion is monotonic along the field line, we can express the cooling rate as a derivative of the colatitude. Therefore,
\begin{equation}
\begin{split}
    \frac{d\gamma_p}{d\theta} &=  \frac{d\gamma_p}{dt} \frac{dt}{ds} \frac{ds}{d\theta}=
    \frac{16\pi}{3} 
    \sigmaeff b \int d\Omega_i \left.\frac{dn}{dx_i d \Omega_i}\right|_{x_i = x_r}  \\
    &\times \RNS \rmax \sin{\theta}\sqrt{1+3\cos^2{\theta}} 
    \begin{bmatrix}
        \mu'_i, \quad p = \perp\\
        \mu'^3_i, \, p = \parallel
    \end{bmatrix}.\\
\end{split}
\label {eqn:dgamma_dtheta}
\end{equation}

We also note that computing the electron cooling via the energy exchange is equivalent to our approach of using the momentum exchange along the $\boldsymbol{B}$ direction. In particular, in the Thomson regime where $x_i'\approx x_f'$, one can prove that $\Delta \gamma =(x_i-x_f) = \beta(x_i\mu_i-x_f\mu_f) = \gamma\beta(x'_i\mu'_i-'x_f\mu'_f)$. 
\subsection{Electron Cooling by the Surface Seed Photons}
\label{sec:surface}

If the magnetosphere is optically thin, the surface thermal emission provides the seed photon for RCS. We assume a Planckian seed photon distribution, %
restricted to the solid angle subtended by the stellar surface as viewed from the scattering site.
The seed photon number density per unit energy and per unit solid angle can be written as a function of incident photon energy and the rescaled dimensionless temperature $\Theta = k_BT/(m_e c^2)$:
\begin{equation}
\begin{split}
    \frac{dn_s}{ dx_i d \Omega_i}
    =
    \neff \frac{ x^2}{e^{x/\Theta}-1}
    ,\quad
    \text{with}\quad 
    \neff\equiv  \frac{\fsc^3}{4\pi^3 r_0^3} .
    \end{split}
    \label{eqn:dn_s_dx_dOmega}
\end{equation}

We further assume the scattering site lies in the far-field regime, sufficiently far from the stellar surface that radiation emitted from different surface elements is approximately parallel. Under this approximation, the angular integration of the incident radiation in Sections \ref{sec:RCS_in_field_loops} and \ref{sec:Electron_Cooling} can be replaced by $\Delta\Omega_i$, the solid angle subtended by the star at the scattering site, which is approximately proportional to $r^{-2}$. Then the cooling relations become
\begin{equation}
    \frac{d\gamma_p}{dt} =\frac{16\pi}{3}\beta c  \sigmaeff \neff b \Delta\Omega_i \frac{x_r^2}{e^{x_r/\Theta}-1} 
    \begin{bmatrix}
         \mu_i' \quad \text{for}\quad p = \perp\\
        \mu_i'^3\; \text{for}\quad p = \,\parallel
    \end{bmatrix},
    \label{eqn:dgamma_dt_surface}
\end{equation}
and
\begin{equation}
\begin{split}
    \frac{d\gamma_p}{d\theta}    =&
    {\frac{16\pi}{3} }
    \sigmaeff \neff b \Delta\Omega_i \frac{x_{r}^2}{e^{x_r/\Theta}-1}  \\
    &\times \RNS \rmax \sin{\theta}\sqrt{1+3\cos^2{\theta}} 
    \begin{bmatrix}
        \mu'_i, \quad p = \perp\\
        \mu'^3_i, \, p = \,\parallel
    \end{bmatrix},
\end{split}
\label{eqn:dgamma_dth_stellar}
\end{equation}
with 
\begin{equation}
    \Delta\Omega_i=2\pi\left(1-\sqrt{1-\frac{1}{\rmax^2\sin^4\theta}}\right).
\end{equation}
Here, $x_r = b/[\gamma(1-\mu_i\beta)]$ is again the resonant energy in the OF at the scattering site. The incident angle can be expressed in terms of the colatitude of the scattering site $\theta$ as 
\begin{equation}
\mu_i=\hat{k}_i\cdot\hat{B}\approx \frac{2\cos{\theta}}{\sqrt{3\cos{\theta}^2+1}}\,\text{sgn}(\pi/2-\theta).
\end{equation}

In the limit of $x_r/\Theta \gg 1$, the cooling rate scales as $\dot{\gamma}\propto -b^2/\gamma$, which is consistent with the {thermal-seed-photon} results in the Thomson limit derived by \cite{1990ApJ...360..197D} and \cite{2011ApJ...733...61B}. {Equation~(\ref{eqn:dgamma_dt_surface}) differs from Equation (B6) of \cite{2013ApJ...777..114B} by a factor of 2 $\times$ 2: one factor of 2 arises from the classical line width, and the other from the Planck spectrum for unpolarized photons.}

As we will show below, the far-field approximation is accurate because near the stellar surface the energies of the surface photons remain far from the resonance, while RCS begins to dominate opacity only at distances greater than a few stellar radii. Note that we do not include RCS by the ultra-relativistic $e^\pm$ particles in the innermost region. The photons produced by such scatterings are expected to convert promptly into pairs and therefore do not contribute directly to the observed spectrum. We account for the results of pair production by absorbing the multiplicity in our electron density normalization.

\begin{figure}
    \centering
    \includegraphics[width=0.9\linewidth]{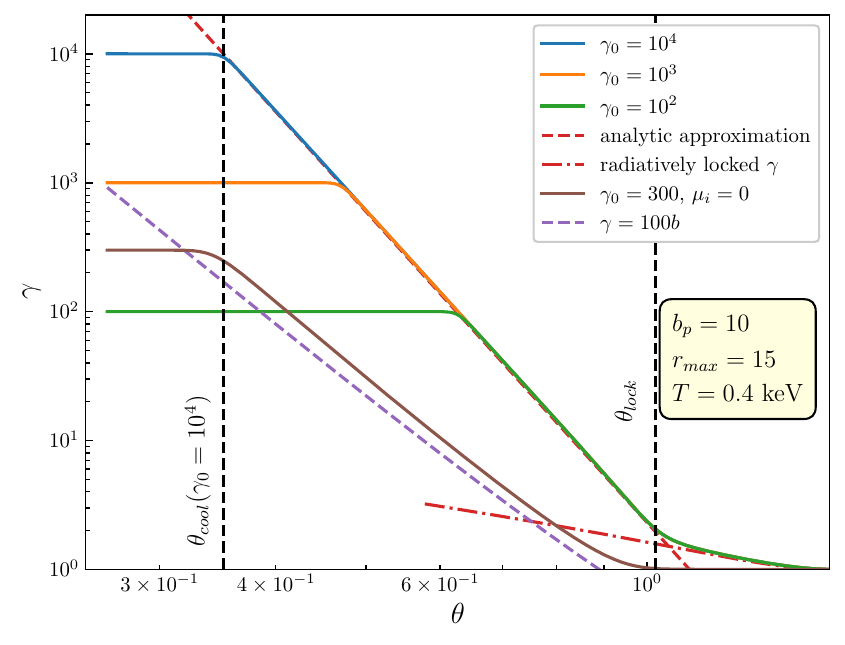}
    \includegraphics[width=0.9\linewidth]{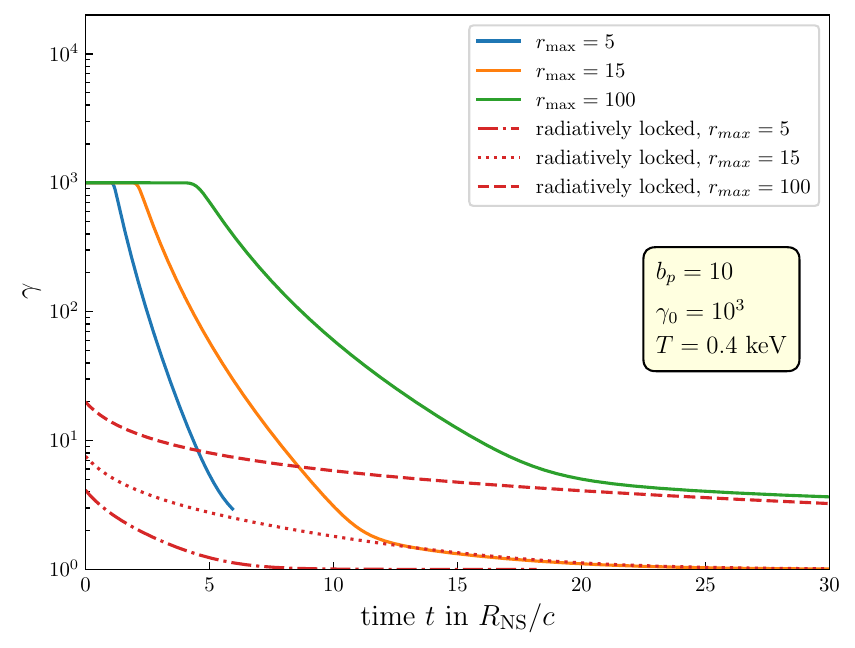}
    \caption{Upper panel: The cooling of electron Lorentz factor $\gamma$ by surface or reflected thermal photons for different initial Lorentz factor $\gamma_0$ as functions of field line colatitudes $\theta$. The analytic power-law approximation in Equation~(\ref{eqn:gamma_theta}) is plotted as dashed red curve and the radiatively locked $\gammalock$ in Equation~(\ref{eqn:gamma_lock}) is plotted as dotted dashed curve. The values of $\thetacool$ and $\thetalock$ are also labeled as vertical dash lines. The analytic approximation of $\gamma=100b$ is also plotted for comparison. Lower panel: Electron Lorentz factor $\gamma$ plotted as functions of time for different field loops labeled by $\rmax$. }
    \label{fig:gamma_vs_theta_b}
\end{figure}

By numerically solving Equation~(\ref{eqn:dgamma_dth_stellar}), we obtain the cooling curve of the electrons along specific field lines. Figure~\ref{fig:gamma_vs_theta_b} shows the electron Lorentz factor $\gamma$ as a function of the magnetic colatitude $\theta$ for different initial Lorentz factors $\gamma_0$ (upper panel), and as a function of time $t$ for different field loops (lower panel).

We find that the cooling rate is highly sensitive to the exponential factor in Equation~(\ref{eqn:dgamma_dth_stellar}), which reflects the Planck form of the seed-photon spectrum. As a result, the cooling curves %
realize three distinct stages that depend on the magnetic colatitude $\theta_{}$. For small $\theta_{}$ %
near the footpoint of the field line, the $\gamma$ curve exhibits a horizontal plateau. In this regime, most seed photons lie outside the resonance, so the RCS cooling is very inefficient, and the electrons stream freely along the field lines. 
{Efficient cooling sets in at a critical colatitude $\thetacool$, where $x_r/\Theta\sim 13$. At this point, the electrons encounter a ``Compton speed bump''\footnote{Arthur H.\ Compton, after whom Compton scattering is named, served as the chancellor of Washington University in St.\ Louis from 1946 to 1954. In 1953, he prototyped a specific type of speed bump and had them installed around the campus to regulate the speed of motor vehicles on the road. These are often considered the predecessors of modern speed bumps. We believe this is a fitting tribute in the context discussed in this paper.}, and their Lorentz factors begin to be regulated by RCS.}
Expanding $x_r$ to the leading order of $\theta$ yields $b\approx b_p/(\rmax^3\theta^6)$ and $x_r\approx 8b_p/(\gamma\rmax^3\theta^8)$. Therefore, we obtain
\begin{equation}
    \thetacool = \left(\frac{8b_p}{13\gamma_0\rmax^3\Theta}\right)^{1/8}.
    \label{eqn:gamma_cool}
\end{equation}
This characteristic angle, $\thetacool$, marks the onset of efficient cooling. %
{Due to the exponential factor $e^{x_r/\Theta}$ in Equation~\eqref{eqn:dgamma_dth_stellar}, the strength of cooling is quite sensitive to the ratio $x_r/\Theta$, and pushes it to roughly remain constant during this rapid cooling stage. As a result, we may write}
\begin{equation}
    \gamma \approx \gamma_0 \left(\frac{\theta}{\thetacool}\right)^{-8}
    =\left(\frac{8b_p}{13\rmax^3\Theta}\right)\theta^{-8}
    \quad \text{for}\quad \theta>\thetacool.
    \label{eqn:gamma_theta}
\end{equation}
Therefore the cooling of $\gamma$ enters a power-law stage as displayed in the upper panel of Figure~\ref{fig:gamma_vs_theta_b}. The normalization and slope of this power-law segment are essentially independent of the initial $\gamma_0$. However, electrons transition into this regime at different $\thetacool$ {depending on their $\gamma_0$, with the dependence given}
by Equation~(\ref{eqn:gamma_cool}).

As electrons stream outward along the field line, their Lorentz factor $\gamma$ can drop rapidly due to radiative losses. Once %
the radiative cooling becomes sufficiently strong, the evolution approaches a saturated “radiatively locked” regime where the radiative force vanishes in the ERF. 
This occurs when $\mu'_i \propto (\mu - \beta)=0$ in Equation~(\ref{eqn:dgamma_dt}), yielding the locked Lorentz factor 
\begin{equation}
    \gammalock = \sqrt{3\cos^2{\theta}+1}/\sin{\theta}
    \label{eqn:gamma_lock}
\end{equation}
\citep[see Appendix B of][]{2013ApJ...777..114B}. Equating Equation~(\ref{eqn:gamma_theta}) with $\gammalock$ determines the corresponding radiatively locked colatitude
\begin{equation}
    \thetalock = \left(\frac{\gamma_0 \thetacool^8}{2}\right)^{1/7}=\left(\frac{4 b_p}{13\rmax^3\Theta}\right)^{1/7}.
    \label{eqn:theta_lock}
\end{equation}
The radiatively locked $\gammalock$ and $\thetalock$ are plotted as a dotted red line and dashed vertical black line in the upper panel of Figure~\ref{fig:gamma_vs_theta_b}, respectively. Again, $\thetalock$ only depends on $\gamma_0$, therefore electrons with the same $\rmax$ all follow the same radiatively locked track. The lower panel of Figure~\ref{fig:gamma_vs_theta_b} displays the electron Lorentz factor $\gamma$ as a function of time, with the radiatively locked case also plotted in red for comparison. As time increases, the numerical cooling curves approach and then follow the
{same radiatively locked evolution.}
For small $\rmax$, however, electrons reach the equator before attaining the radiatively locked state. If an electron is emitted with a very small $\gamma$, it initially moves horizontally in the $\gamma$--$\theta$ plane until the resonance condition is satisfied. If the electron enters the resonance below the radiatively locked curve, it is driven toward that curve because the net radiation force is directed outward in the ERF.

{Recently, the radiatively locked state was invoked to explain the observed radio emission from magnetars~\cite{2026ApJ...996L..20Z}. The key observation is that electrons and positrons require a small velocity separation in order to conduct the magnetospheric current. This velocity separation tends to trigger the two-stream instability, leading to a thermal spread in the momentum distributions of both particle species which eventually halts the instability growth. However, the radiation force keeps the momentum distributions narrow, leading to continual excitation of the two-stream instability and sourcing coherent radio emission.}

Assuming $\thetacool\sim \pi/2$, Equation~(\ref{eqn:gamma_cool}) implies that electrons can remain out of resonance essentially all the way to the magnetic equator if 
\begin{equation}
    \rmax \lesssim 0.26\left(\frac{b_p}{\gamma_0  \Theta} \right)^{1/3}\quad.
\end{equation}
Similarly, from Equation~(\ref{eqn:gamma_theta}) one finds that RCS is sufficient to cool the electron to $\gamma \sim1$ if
\begin{equation}
    \rmax \gtrsim 0.26\left(\frac{b_p}{ \Theta} \right)^{1/3}
    \approx 5.5\left(\frac{b_p}{10}\right)^{1/3}\left(\frac{\Theta}{0.5\,\text{keV}/m_ec^2}\right)^{-1/3}\quad.
    \label{eqn:rmax_cool}
\end{equation}
Equation~(\ref{eqn:rmax_cool}) suggests that for a typical magnetar with $b_p\sim10$ and a surface temperature $kT\sim0.5$ keV, the RCS cooling by the surface thermal photons is strong enough to lock the electrons before they reach the magnetic equator for field loops with $\rmax>5.5$.  
In this case, electrons are efficiently cooled and stall near the equator, and cross-hemisphere streaming is prevented. While inside the $\rmax\sim5.5$ region, electrons can stream across the equatorial region and enter the opposite hemisphere.

\subsection{Electron Cooling by Reflected Photons}
\label{sec:cooling-thick}

If the magnetosphere is optically thick to the surface seed photons, the thermal photons may be scattered multiple times before being resonantly upscattered to high energies. {Since electrons enter the radiatively locked state close to the equator, they can build up a large density and become optically thick to the thermal photons from the star. This effect has been predicted by \citet{2013ApJ...777..114B}. In this limit, the seed photons for RCS may come from a large region in the magnetosphere, in addition to the stellar surface.} {We consider the extreme case where the entire magnetosphere is optically thick, and the seed photons for RCS are essentially isotropic, $\langle \mu_i\rangle \approx 0$ in the OF.}

{Assuming that the seed photon spectrum is not strongly modified by the optically thick scattering in the magnetosphere, the exponential factor $e^{x_r/\Theta}$ still enters $d\gamma_p/d\theta$ in the same way, which regulates the ratio $x_r/\Theta$ to remain roughly constant, $x_r/\Theta \sim 13$. This constant will depend on the density of the isotropic photon field, so we take the same value as in Section~\ref{sec:surface}. Such a radiation regulated ratio implies a scaling relation of:}
\begin{equation}
\gamma \approx \frac{b}{13\Theta} \approx 77 \left(\frac{\Theta}{10^{-3}m_e c^2}\right)^{-1}b\propto \theta^{-6}.
\label{eqn:gamma_propto_100b}
\end{equation}
 We note Equation~(\ref{eqn:gamma_propto_100b}) closely resembles the {empirical relation} $\gammaSC\approx (m_ec^2/10kT)b\approx100b$ in Equation~(37) of \cite{2013ApJ...777..114B}, %
 which assumes $\cos{\theta_i}\sim-0.5$. This relation is also supported by their Monte Carlo simulations, which approximate a self-consistent description of an optically thick plasma outflow in magnetar magnetospheres. 
 Note that the radiatively locked stage does not exist for the reflected seed photon case, since $\mu_i'\propto(\mu_i-\beta)$ is always negative in Equation~(\ref{eqn:dgamma_dt}).

 In the upper panel of Figure~\ref{fig:gamma_vs_theta_b}, we also plot the $\gamma = 100b$ curve and the reflected cooling curve obtained by setting $\mu_i=0$ in Equation~(\ref{eqn:dgamma_dth_stellar}). After entering the resonance, the reflected cooling curve is very similar to $\gamma = 100b$, with a slight deviation that is likely attributable to uncertainty in the scaling factor. If the initial Lorentz factor $\gamma_0$ exceeds $100 b_p$, the $e^\pm$ particles enter the resonance immediately, causing the cooling curve to drop rapidly and approach the $\gamma = 100b$ curve at relatively small $\theta$. It can also be seen from the figure that the plasma does not enter a radiatively locked state.

The simple algebraic scaling in Equation~\eqref{eqn:gamma_propto_100b} breaks down far away from the star when $b$ becomes very small. Naively applying the equation would predict a Lorentz factor $\gamma < 1$, which is unphysical. This signals that the isotropic approximation $\langle \mu_i\rangle \approx 0$ breaks down, and that other physics, including plasma kinetics and pair annihilation, will likely become important. Moreover, the $\mu_i$-averaged energy-loss formula breaks down at very high $\gamma$, because an $e^\pm$ particle can lose most of its energy in a single scattering event. A fully self-consistent calculation that treats the plasma physics and radiation transport simultaneously will be needed in this regime, which lies beyond the scope of the present paper.

\begin{figure*}
    \centering
    \includegraphics[width=0.49\linewidth]{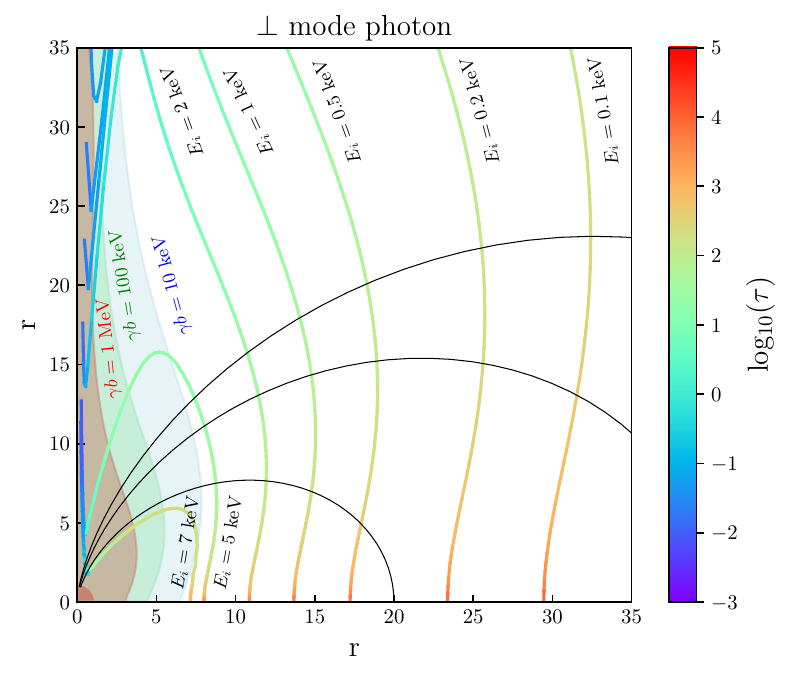}
        \includegraphics[width=0.49\linewidth]{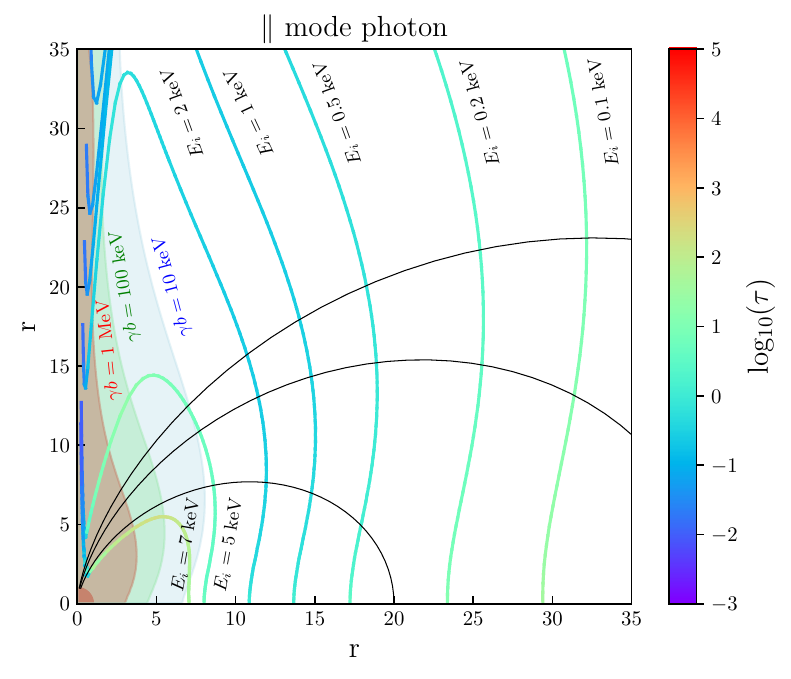}
    \caption{Location of resonance for $\perp$ (left panel) and $\parallel$ (right panel) mode photons radially emitted from the surface for a star with $b_p=10$ and $\Theta = 0.5$ keV. The electron density at surface $n_0$ is assumed to be $10^{18}$ cm$^{-3}$. The red circle at the origin represents the size of the neutron star. Selected dipole magnetic field lines with $\rmax=$ 20, 40, and 60 are also plotted for comparison. The color coding in \red{both panels} represents the optical depth a photon achieves when crossing the resonant location. The shaded area represents the region where $\gamma b$ is greater than the selected energies of 10 keV, 100 keV and 1 MeV. %
    } 
    \label{fig:r_reso_tau}
\end{figure*}
%
\begin{figure*}
    \centering
    \includegraphics[width=0.49\linewidth]{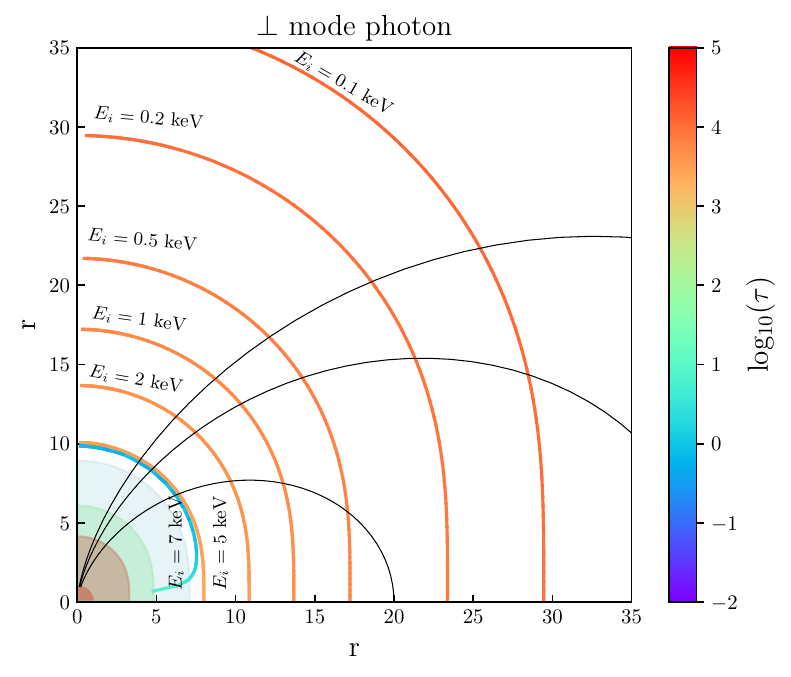}
    \includegraphics[width=0.49\linewidth]{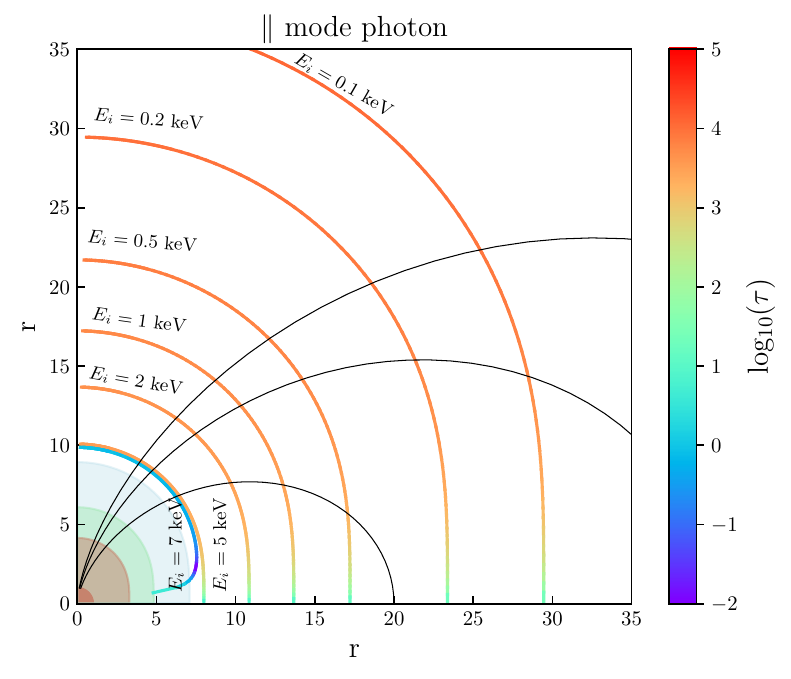}
    \caption{Similar to Figure. \ref{fig:r_reso_tau} but using $\gamma = 100b$ for the electron Lorentz factor. The electron density at surface $n_0$ is assumed to be $10^{16}$ cm$^{-3}$. Again, the shaded area represents the region where $\gamma b$ is greater than 10 keV, 100 keV and 1 MeV. %
    We set $\gamma$ to 1 for region with $b < 0.01$ and impose an upper limit $\nmax\sim 5\times10^{22}$ cm$^{-3}$ to prevent divergence. 
    }
    \label{fig:r_reso_tau_reflect}
\end{figure*}

\subsection{{Magnetosphere Opacity for Surface X-ray Photons}}
\label{sec:opacity}
Using the cooling curves calculated in the last section, we can now compute the optical depth experienced by a thermal X-ray photon propagating radially outward from the stellar surface. {This calculation is necessary to ensure \emph{self-consistency}, i.e.\ if the cooling curve assumes the optically thin limit, then the computed optical depth should be less than unity, and vice versa.}

We assume that the electron Lorentz factor $\gamma$ evolves according to the cooling calculation, %
and that the electron density $n_e$ satisfies the charge continuity condition given in Equation~(\ref{eqn:n_e}). The differential optical depth can be written as {(assuming the photon originates from the star and propagates radially outwards):}
\begin{equation}
    \begin{split}
    \frac{d \tau}{dr}&=n_e (1-\beta\mu_i)\int dx'_fd\Omega'_f \frac{d\sigma'}{dx' d\Omega'}\\
    &\approx\frac{4\pi^2 r^2_0}{\fsc} n_e (1-\beta\mu_i)
    \begin{bmatrix}
        1, \,\text{for} \perp \,\text{photon}\\
        \mu_i'^2, \,\text{for} \parallel \,\text{photon}
    \end{bmatrix}\delta(x'-b)\quad.
\end{split}
\label{eqn:dtau_ds}
\end{equation}
Again, $x' = x\gamma(1-\mu_i\beta)$ is the incident photon energy in the ERF. Since $\gamma$ is a function of location and $\mu_i$ is a constant for radially emitted photons, $x'$ can be expressed as a function of the photon radius $r$.
The $\delta$ function vanishes after the integration and Equation~(\ref{eqn:dtau_ds}) yields
\begin{equation}
    \tau = \frac{4\pi^2 r^2_0}{\fsc} 
    \frac{n_e (1-\beta\mu_i)}{|dx'/dr+3b/r|}
    \begin{bmatrix}
        1, \,\text{for} \perp \,\text{photon} \\
        \mu_i'^2, \,\text{for} \parallel \,\text{photon}
    \end{bmatrix}H(r-\rres)
    \label{eqn:tau}
\end{equation}
with $H(r-\rres)$ being the Heaviside step function. Here $\rres$ is the radius at which the resonant condition $x'=b$ is satisfied. 
\red{The optical depth calculated here represents the probability of the initial scattering for a photon traversing the magnetosphere from the stellar surface. It should not be broadly interpreted as determining the mean number of scatterings experienced by the photon, because the photon energy and momentum, as well as the differential optical depth $d\tau$, change drastically after the first scattering event. This distinction is fundamentally tied to the kinematics of the electrons. For RCS by relativistic electrons, the initial scattering tightly beams the outgoing photon along the electron's velocity vector (the local magnetic field direction), such that $d\tau$ and the collision rate for any subsequent scatterings are significantly suppressed through the $(1-\beta\mu_i)$ factor. Conversely, for scattering by non-relativistic electrons in the equatorial region, the extreme relativistic beaming is absent. In that regime, a high initial optical depth causes photons to become trapped in the semi-transparent resonant layer \citep[see][]{2006MNRAS.368..690L}, and the calculated $\tau$ does indeed reflect the expected number of scatterings.
}

In Figure \ref{fig:r_reso_tau} we display the resonant radius $\rres$ in the magnetosphere for a selection of radial seed photon energies $E_i$ for both polarization modes. The color scale encodes the optical depth calculated using Equation~(\ref{eqn:tau}), assuming that the whole magnetosphere is twisted and filled with charges. Due to the Heaviside function in Equation~\eqref{eqn:tau}, the optical depth quickly jumps from zero to the maximum value along a given line of sight and remains constant afterwards, so it is sufficient to plot this maximum value at the resonant surface. We fix the normalization of the number density $n_0$ to 10$^{18}$ cm$^{-3}$, which is a representative number that yields model spectra with luminosities comparable to observed values (see Section~\ref{sec:spectrum}). In general, %
lower-energy photons resonate at larger radii where the field strength is weaker, thereby producing more extended resonant contours. 
The $\perp$ and $\parallel$ mode photons yield broadly similar contours, with slight deviations caused by the small difference in the electron cooling for the two modes.

An interesting feature of Figure~\ref{fig:r_reso_tau} is that a substantial fraction of the magnetosphere, measured over emission directions relative to the magnetic axis, can be optically thick to the surface $\perp$ mode photons. %
If the whole magnetosphere is filled with charges, %
transparency for the $\perp$ mode is achieved only for photons emitted within a 
{rather small}
angular range about the magnetic axis. %
It can also be seen that at a given photon energy $E_i$, the optical depth $\tau$ for the $\perp$ mode seed photons increases with the emission colatitude $\theta$, transitioning from the optically thin regime ($\tau \ll 1$) in the polar region to the optically thick regime ($\tau \gg 1$) in the equatorial region. This feature was also noticed by~\citet{2013ApJ...777..114B}. The rise in optical depth with increasing emission colatitude can be understood as the combined effect of plasma accumulation and kinematic weighting in the scattering coefficient. As the outflow cools toward the equatorial region, the plasma consequently builds up and the number density increases. In addition, the $(1-\beta\mu_i)$ factor in Equation~(\ref{eqn:tau}) increases toward larger colatitudes, further enhancing the RCS optical depth.
Note that, although the magnetosphere can become optically thick at large $\theta$, our cooling calculations using the unattenuated Planck seed photon spectrum are not affected. This is because photons of different energies satisfy the resonance condition at different radii.

For the $\parallel$ mode photons, by contrast, the magnetosphere is largely transparent in most directions, and becomes optically thick only for higher-energy photons that resonate at lower altitudes. The difference in RCS opacity between the two polarization modes is driven mainly by the $\mu'^2_i$ factor in Equation~(\ref{eqn:tau}). Along a given field loop, the pair plasma is ultra-relativistic at small magnetic colatitude $\theta$. In this regime, strong relativistic beaming yields $\mu'^2_i\sim1$, so the opacities for the two polarization modes become effectively the same. At larger $\theta$ or near the equatorial region, however, the plasma becomes only mildly relativistic or even non-relativistic. Consequently, the $\mu'^2_i$ factor can significantly reduce the opacity for $\parallel$ mode photons, making the magnetosphere effectively transparent for the $\parallel$ mode even in the equatorial region.

In Figure~\ref{fig:r_reso_tau}, we also shade the region where $\gamma b$ exceeds 10 keV, 100 keV, and 1 MeV. Because the scattered photon energy satisfies $x_f = \gamma b(1+\mu'_f\beta)\sim\gamma b$, the shaded areas represent the portion of the magnetosphere inside which RCS can upscatter photons to the corresponding labeled energies. This construction is analogous to the ``Compton resonasphere" introduced in \cite{2007Ap&SS.308..109B}, however, here we display the outgoing photon energies directly from our cooling calculation, instead of characterizing the region in terms of the resonant parameter $\Psi=b/(2\gamma x_i)$.

The shaded resonant regions indicate that, the RCS can produce hard X-rays ($>$10 keV) only inside a few stellar radii when the seed photons are emitted in near-equatorial directions. In contrast, if the seed photons are emitted closer to the magnetic-axis direction, the resonant region can extend to much higher altitudes in the polar zone. The extent of this polar resonance is primarily set by the initial Lorentz factor $\gamma_0$, with larger $\gamma_0$ producing a more extended resonant region. Note that, even though the overall opacity for soft X-ray photons is much lower for the $\parallel$ mode, the opacities in the shaded regions are actually comparable between $\perp$ and $\parallel$ mode photons. This indicates that the hard X-ray emission alone may not be enough to differentiate between these two types of seed photons. As can be seen in Section~\ref{sec:spectrum}, the two seed photon modes indeed produce almost indistinguishable hard X-ray spectra.

{To compare with the results of Figure~\ref{fig:r_reso_tau}, we also consider the case of RCS cooling by reflected photons that are isotropic in the entire magnetosphere. There are several key differences. Firstly, RCS cooling by reflected photons is much more efficient at reducing the electron energy since it is much more likely to experience a head-on collision. 
This results in an overall larger plasma density and optical depth in the magnetosphere.
Additionally, since the photons are isotropic along every field line, even electrons flowing out near the pole can be efficiently cooled by RCS. Figure~\ref{fig:r_reso_tau_reflect} shows the resonance surfaces and corresponding optical depths for $\perp$ and $\parallel$ photons assuming an isotropic reflected background photon field. Even at a much lower density normalization $n_0$, the entire magnetosphere is highly opaque to ${\sim}\mathrm{keV}$ photons originating from the star. The only exception is that $\parallel$ mode photons may be able to escape the magnetosphere near the magnetic equator, since both the $\parallel \to \perp$ and $\parallel \to \parallel$ cross sections depend on $\cos^2\theta_i$ which is approximately zero along the equator.}

\subsection{Is the magnetosphere optically thick?}
\label{sec:optical-thick}

Interpretations of magnetar soft X-ray emission exclusively focus on the scenario in which the soft X-rays originate from the stellar surface \citep[][]{2001MNRAS.327.1081H,2001ApJ...563..276O,2001ApJ...560..384Z}. This is strongly justified since spectral fitting often reveals an emission site size of typically $1$--$5\,{\rm km}$ \citep[][]{2007MNRAS.381..293R,2008ApJ...680L.133T}, and the typical sinusoidal pulse profiles suggest that the radiation comes from a hotspot.
In these studies, a magnetized atmosphere is assumed to cover the surface of magnetars, and the surface thermal emission is expected to be dominated by the $\perp$ mode photons, because their opacity is strongly suppressed relative to values for the $\parallel$ mode photons.
However, the results outlined in Section~\ref{sec:opacity} highlight the fact that, if the entire magnetar magnetosphere is twisted, then to support the observed hard X-ray flux, the electron number density needs to be high enough that most of the magnetosphere is opaque to soft X-ray $\perp$ mode photons emitted from the star.
To reconcile this apparent transparency with the requirement of bright hard X-ray emission, we propose two potential ways to avoid a completely optically thick magnetosphere.

{According to Figure~\ref{fig:r_reso_tau}, soft X-ray photons between $0.5\,\mathrm{keV}$ and $5\,\mathrm{keV}$ originating from the star will only resonantly scatter beyond $r \sim 8R_*$. Therefore, if the twisted field line bundle lies completely within this radius, then most of the soft photons will be able to escape the magnetosphere without resonantly scattering with the plasma. On the other hand, there is also a finite range of angles near the magnetic poles where the resonant surface becomes further away from the star and the scattering optical depth drops below unity. Therefore, if the twist is concentrated close to the polar cap, or small $\theta$, the plasma flow will not strongly scatter the soft X-ray photons from the surface. The latter scenario was also what was considered by~\citet{2014ApJ...786L...1H} when fitting the \emph{NuSTAR} spectra of select persistent magnetars. The two configurations are shown in Figure~\ref{fig:geometry}. In the rest of this paper, we focus our attention on these two scenarios, and attempt to compute the resulting hard X-ray spectra due to resonant inverse-Compton scattering.}

\section{RCS Hard X-ray Spectrum}
\label{sec:spectrum}
{In this section, we will first outline how we compute the hard X-ray emission spectrum from the plasma properties and the assumed twist profile. Then, we perform the calculation for the two potential scenarios proposed in Section~\ref{sec:optical-thick} where most of the magnetosphere remains optically thin to soft X-ray photons from the star. In these calculations, we use the magnetic inclination angle and observer angle inferred from fitting a rotating vector model (RVM) to the \emph{IXPE} polarization measurements of one particular persistent magnetar 4U 0142+61, for which the RVM seems to be a good description. This helps tremendously in reducing the allowable parameter space, and allows us to roughly assess whether these two scenarios are indeed compatible with the observed phase-resolved hard X-ray spectra seen by \emph{NuSTAR}.}

\subsection{Hard X-ray Spectrum Calculation}

\begin{figure}
    \centering
    \includegraphics[width=0.95\linewidth]{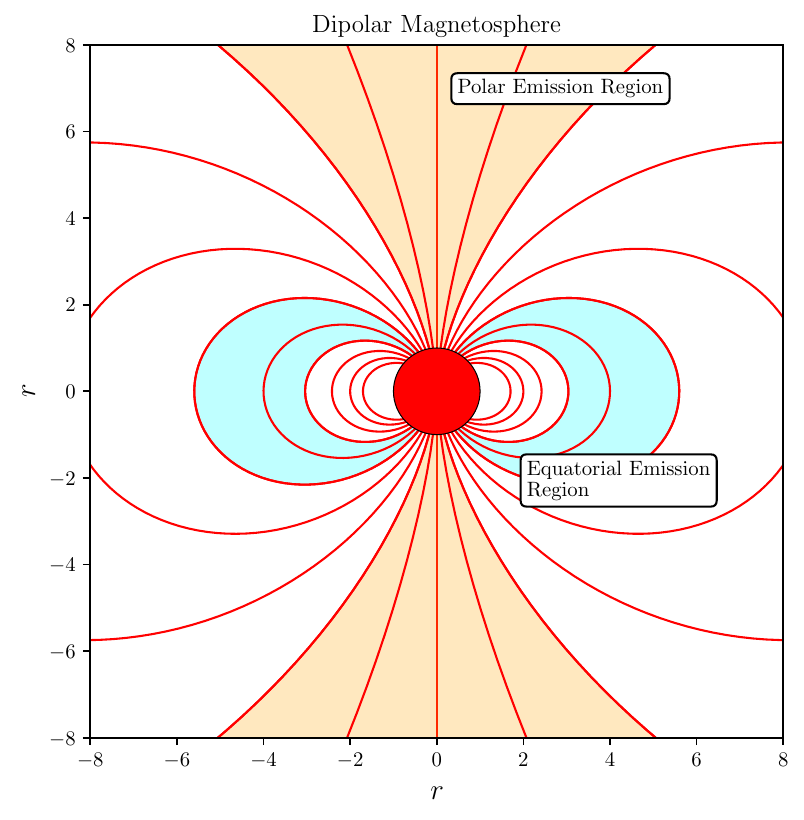}
    \caption{Geometry of the axisymmetric emission regions. We consider two configurations: (1) a polar region defined by either the foot point colatitude $\theta_f$ or by an outer field-line boundary labeled by $\rmax$; and (2) an equatorial region bounded by two sets of field lines.}
    \label{fig:geometry}
\end{figure}

To move beyond RCS on a single field line with constant electron $\gamma$ factor, we numerically integrated Equation~\eqref{eqn:dNph_dt_dxf_dOf_drmax_dtheta} over $\rmax$ and $\theta$ (or Equation~\eqref{eqn:dNph_dt_dxf_dOf_drmax_dtheta_2} if $\beta$ is small). 
We also assume the thermal distribution of the seed photon in Equation~(\ref{eqn:dn_s_dx_dOmega}) and adopt the far-field approximation to replace $\int d\Omega_i$ by $\Delta\Omega_i$. Then the RCS spectrum in Equation~(\ref{eqn:dNph_dt_dxf_dOf_drmax_dtheta}) can be written as
\begin{equation}
\begin{split}
    \frac{dN_{p \rightarrow q}}{dt\, dx_f\,   
    d\Omega_f d\rmax d\theta}&(x_f,\rmax,\theta,\Omega_f) \\%
    =\sum_{\phires}
    n_e \rmax^2\sin^7{\theta} & \RNS^3 \frac{c \Delta \Omega_i}{\gamma^2 (1-\beta \cos{\theta_f})}\\
    \times &\sigmaeff \neff \frac{x_r^2}{e^{x_r/\Theta}-1}
    \frac{{\cal P}_{p\rightarrow q}}{|g'(\phi)|}.
\end{split}
\label{eqn:dNph_dt_dxf_dOf_drmax_dtheta_far_field}
\end{equation}

For a given choice of $\rmax$, the integration range of $\theta$ is set by the footpoint colatitude of the corresponding field line. The evolution of $\gamma$ along the field loop is calculated using Equation~(\ref{eqn:dgamma_dtheta}), while the electron density follows Equation~(\ref{eqn:n_e}) with $n_0$ treated as a free normalization parameter. We impose a density cap $\nmax = 5 \times10^{22}$ cm$^{-3}$ everywhere to avoid divergence in $n_e$ {when the plasma decelerates to a halt near the equator}. %
We have verified that the resulting hard X-ray spectra are insensitive to the exact value of $\nmax$.

We consider two types of axisymmetric scattering regions, each bounded by selected magnetic field lines (see Figure~\ref{fig:geometry}). 
The first is an equatorial region, defined by an inner and outer boundary in $\rmax$. This configuration closely resembles a localized twist confined to the closed-field-line zone. The second is a polar-cap region, specified either by the footpoint colatitude of the bounding field line or, equivalently, by its maximum field line $\rmax$. We also truncate the scattering region at a maximum radius of $60\RNS$; above that radius the magnetic field is weak and RCS is inefficient in producing hard X-rays. {Inside the twisted region, we take the density normalization $n_0$ to be the same across field lines, which roughly translates to a uniform current density profile. Realistic current profiles require the solution of the force-free equilibrium in the given twist configuration, which goes beyond the scope of this paper, and will be incorporated in future work.}

\subsection{Hard X-ray Spectrum from Equatorial Magnetospheric Regions}
\label{sec:equatorial_spec}

\begin{figure}
    \centering
    \includegraphics[width=0.95\linewidth]{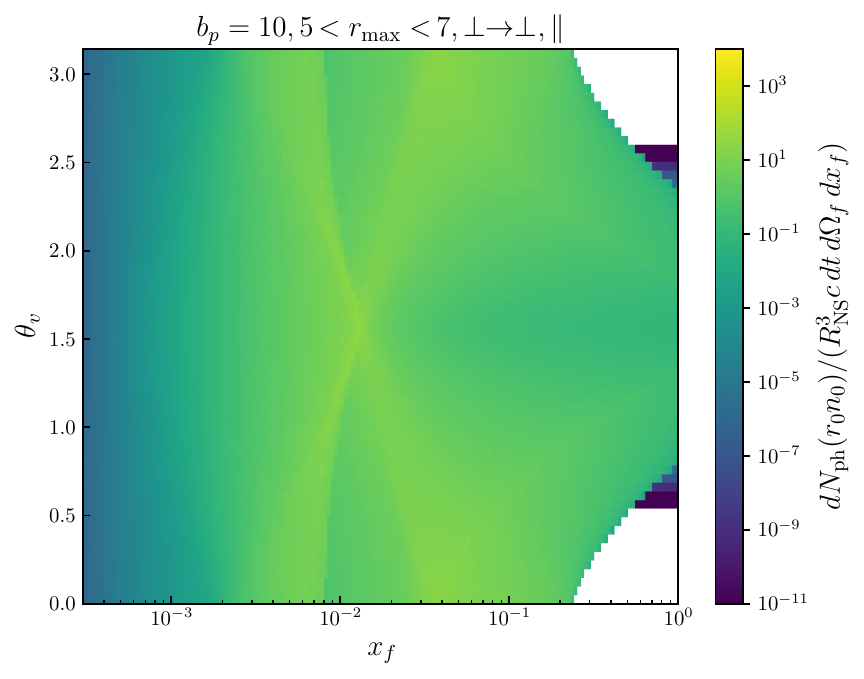}
    \caption{Calculated RCS photon number flux plotted as maps of the final photon energy $x_f$ (in units of $m_ec^2$) and viewing angle $\theta_v$ with respect to the magnetic axis. The polar field strength is $b_p = 10$ and the initial electron density $n_0$ is set to $5\times 10^{18}$ cm$^{-3}$. 
    The flux map is calculated assuming thermal surface seed photons with temperature $kT = 0.4$ keV. %
    }
    \label{fig:thetav_xf_map}
\end{figure}

{First, we consider the equatorial twist profile where the active flux bundle is within some small range of $r_{\rm max}$.}
In Figure~\ref{fig:thetav_xf_map}, we display the RCS photon number flux calculated as heat maps of final photon energy $x_f$ and viewing angle $\theta_v$ with respect to the magnetic axis. The seed photons are assumed to come from the stellar surface with a blackbody temperature $kT = 0.4$ keV. The flux map is integrated over the inner magnetospheric region $5<\rmax<7$ outside the star. The integration over the azimuthal direction is properly treated using Equation~(\ref{eqn:dNph_dt_dxf_dOf_drmax_dtheta_far_field}). For a given viewing angle $\theta_v$, the RCS spectrum starts from very low energy and extends above a few hundred keV. Higher energies can be achieved at larger $\theta_v$, where the line of sight is tangent to the field direction at smaller altitude and larger field strength.

Horizontal cuts of Figure~\ref{fig:thetav_xf_map} give photon number spectra at specific instantaneous viewing angles $\theta_v$. Given the magnetic field inclination $\alpha$ and the angle $\beta$ between the rotation axis and the line of sight, the instantaneous viewing angle $\theta_v$ can be related to the rotation phase as
\begin{equation}
    \cos{\theta_{v}}=\sin{\alpha}\sin{\beta}\cos{\phi}+\cos{\alpha}\cos{\beta}.
    \label{eqn:rotation_phase}
\end{equation}
Here we obtain the phase-resolved spectra with fixed $\alpha \approx 0.25$ and $\beta \approx 1.0$. These geometric angles are approximated from the polarization angle curve of Figure 3 for 4U 0142+61 in \cite{2022Sci...378..646T}, using the rotating vector model \citep[][]{1969ApL.....3..225R}. %

\begin{figure}
    \centering
    \includegraphics[width=0.9\linewidth]{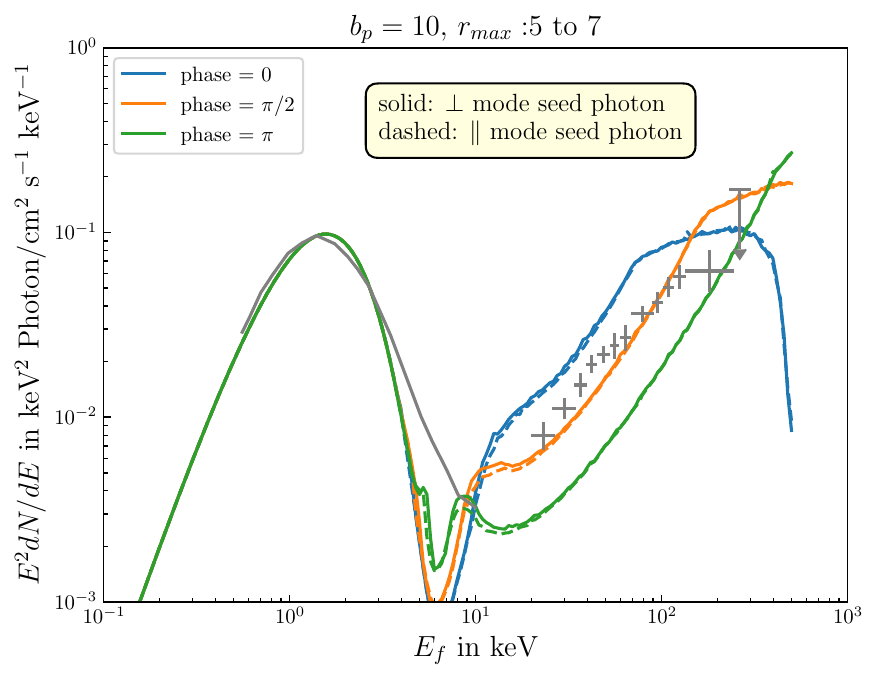}
    \caption{Calculated phase-resolved spectra for the cases where seed photons are from the stellar surface. %
    Different colors represent different rotational phases, and the solid and dashed curves display the spectra for different polarization modes of the seed photon. We assume the surface photons have a temperature $kT=0.4$ keV, the initial Lorentz factor $\gamma_0 = 10^4$, and we rescale the initial number density $n_0$ to $3.2\times 10^{18}$ cm$^{-3}$. The grey data  (including the grey thermal component) are the phase-averaged spectrum of 4U 0142+61 taken from Figure~3 of \cite{2008A&A...489..245D}.}
    \label{fig:phase_resol_spectra}
\end{figure}
\begin{figure}
    \centering
    \includegraphics[width=0.9\linewidth]{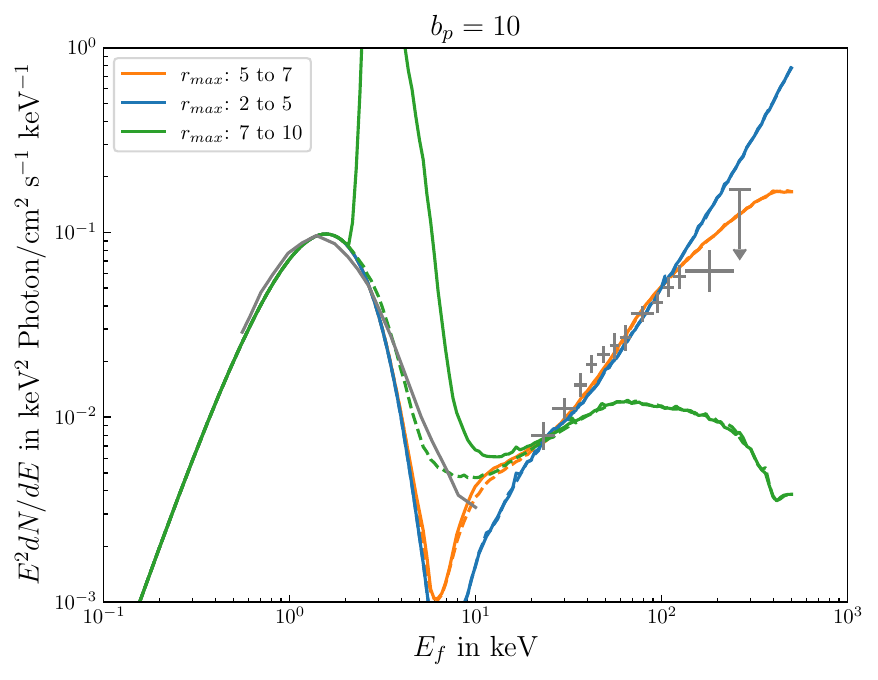}
    \caption{Rotational phase averaged spectra for the regions with %
    $2<\rmax<5$, $5<\rmax<7$, and $7<\rmax<10$.
    Again, the solid curves are for $\perp$ mode seed photons and the dashed curves are for $\parallel$ mode. }
    \label{fig:phase_ave_spectra}
\end{figure}

Figure~\ref{fig:phase_resol_spectra} displays the phase-resolved spectra %
derived from the flux heat map using Equation~(\ref{eqn:rotation_phase}). 
For comparison with the observations, we add a phase-independent blackbody component with $kT = 0.4$ keV to the calculated RCS spectrum. This component is assumed to originate uniformly from the stellar surface for the purpose of our calculation, therefore it does not modulate with rotation phase.
The electron density $n_0$ at the stellar surface 
is chosen so that the phase-averaged spectrum reproduces the X-ray flux observed by \emph{NuSTAR} at 20 keV (see Figure~\ref{fig:phase_ave_spectra}). 
Hereafter, solid and dashed curves denote results for the $\perp$ and $\parallel$ modes seed photons, respectively.
Although the underlying emission region is identical, %
the phase-resolved spectra exhibit substantial variations in both flux level and cutoff energy. This diversity arises because %
different rotational phases sample different instantaneous viewing colatitudes $\theta_v$. 
We also find that phases with lower cutoff energies typically correspond to higher fluxes, which is consistent with the trend seen in  Figure~\ref{fig:thetav_xf_map}.

In our calculations, %
the two polarization cases differ only slightly at $E_f\lesssim$ a few tens of keV, and the discrepancy becomes negligible at higher energies. 
This behavior is expected since higher scattered energies require larger electron Lorentz factors. At large Lorentz factor the scattering is strongly beamed and approaches head-on scattering in the ERF. In this case the polarization dependence becomes weak and the two modes effectively converge.

The phase-resolved spectra in Figure~\ref{fig:phase_resol_spectra} can be integrated over the rotational phase to obtain phase-averaged spectra. Figure~\ref{fig:phase_ave_spectra} shows the resulting spectra for %
three scattering volumes, $2<\rmax<5$, $5<\rmax<7$, and $7<\rmax<10$, 
plotted as blue, orange, and green curves, respectively. In general, 
twists confined to the inner magnetosphere produce spectra with a stronger high-energy component, because the magnetic field strength and the resonant energy are higher at smaller radii. 
The $2<\rmax<5$ and $5<\rmax<7$ cases produce relatively flat power-law hard X-ray tails extending beyond a few hundred keV, which broadly resemble the observed hard X-ray emission of 4U 0142+61. In both cases, however, our calculations underpredict the observed soft X-ray flux which is often attributed to a steep power-law component produced by repeated scattering or to an additional hotter thermal component. 
By contrast, the outer region $7<\rmax<10$ case produces a softer hard X-ray component that turns over at around 100 keV. %
For more extended scattering regions, the hard X-ray power-law component hardens again due to the shift in the dominant emission site of the hard X-ray photons. However, the $e^\pm$ particles cool efficiently in the extended regions, making the equatorial scattering region optically thick. This breaks the self-consistency of the calculations of the plasma cooling and RCS spectra in the optically thin regime and should not be interpreted as a physical prediction.

\subsection{Hard X-ray Spectrum from Polar Magnetospheric Regions}
\label{sec:polar-twist-spec}
\begin{figure}
    \centering
    \includegraphics[width=0.95\linewidth]{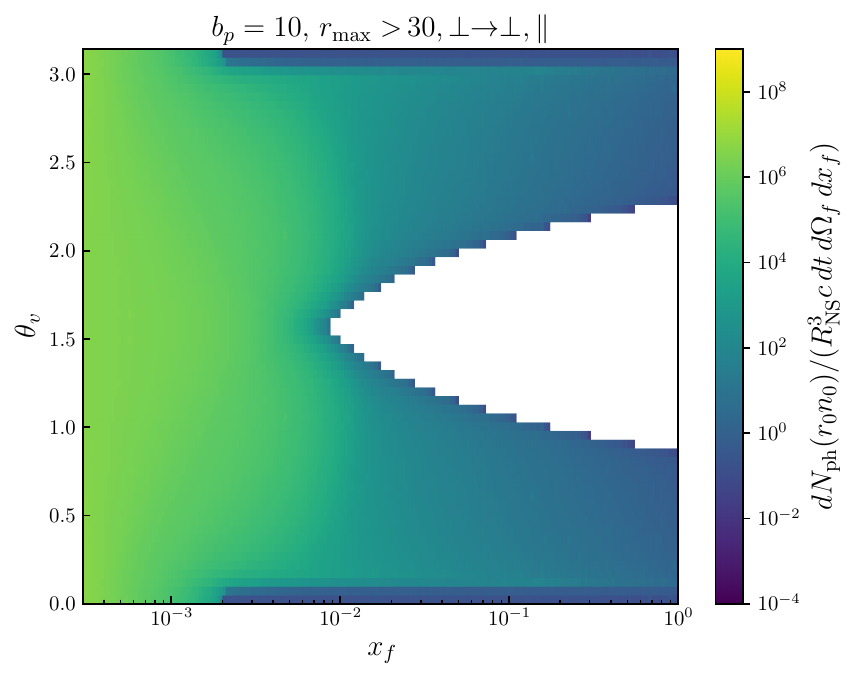}
    \caption{RCS photon number flux plotted as maps of the final photon energy $x_f$ and viewing angle $\theta_v$. The flux is integrated over 
    $\theta_{\rm foot}<{0.18}$ ($\rmax>{30}$), and $r< 60$. }
    \label{fig:thetav_xf_map_outer}
\end{figure}
%
\begin{figure}
    \centering
    \includegraphics[width=0.95\linewidth]{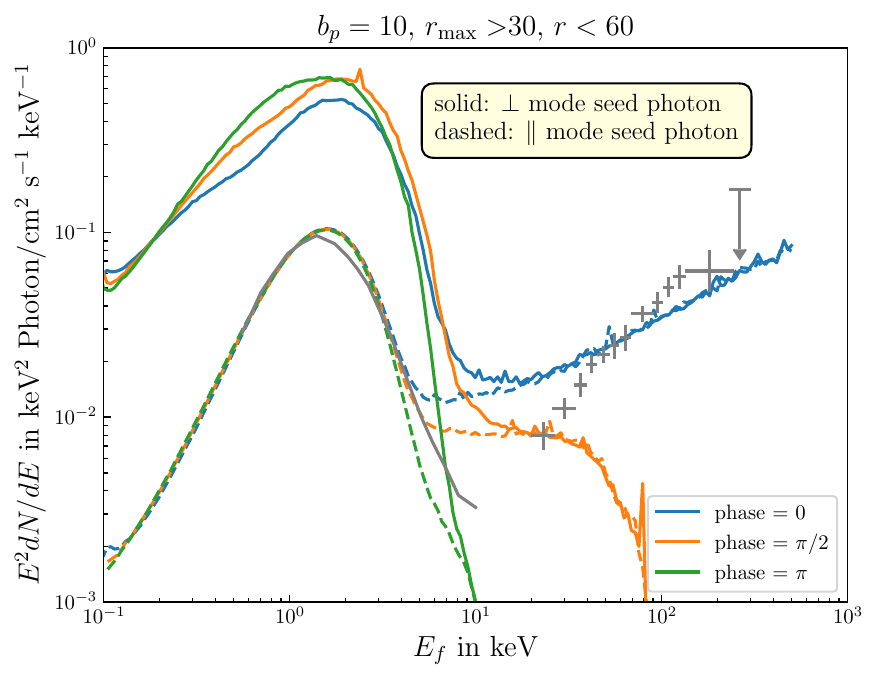}
    \caption{Phase-resolved spectrum for the polar emission region above $\rmax>30$, for the surface seed photon case with $\perp$ (solid) or $\parallel$ (dashed) mode polarization.}
    \label{fig:phase_resol_spectra_rm_gt_30}
\end{figure}
\begin{figure}
    \centering
    \includegraphics[width=0.95\linewidth]{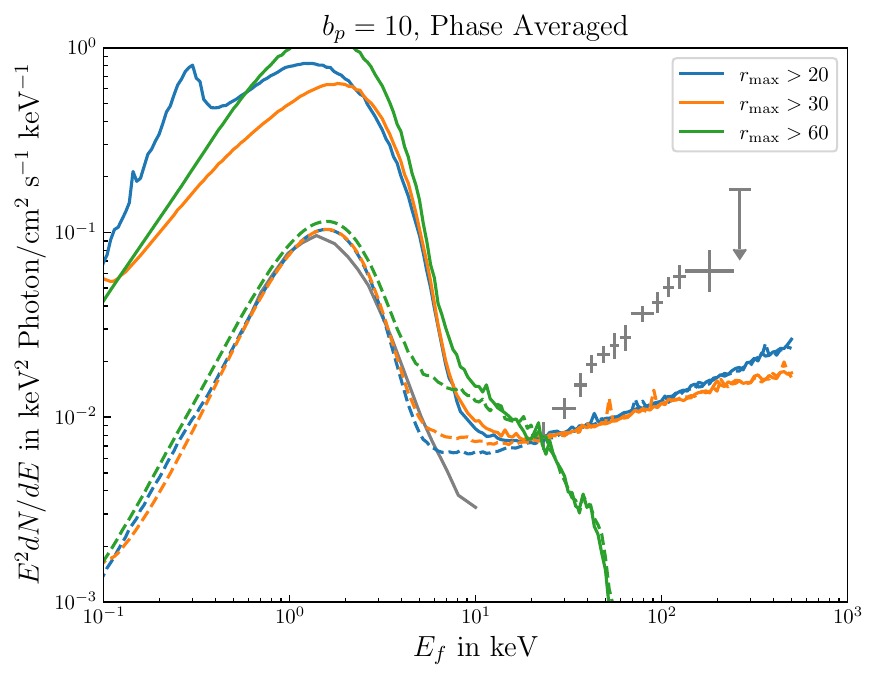}
    \caption{Phase-averaged spectrum for the polar emission regions with $\rmax>20$, $\rmax>30$, and $\rmax>60$. The solid and dashed curves are for $\perp$ and $\parallel$ mode, respectively.}
    \label{fig:phase_ave_spectra_rm_gt_20_30_60}
\end{figure}
In the previous subsection, we examined the integrated RCS spectra for charges confined within an axisymmetric flux tube specified by a range of the $\rmax$ values in the equatorial region. %
The same framework can be extended to larger $\rmax$ to probe emission from the polar region. {As a demonstration, we consider a polar twist that contains all magnetic field lines with $r_{\rm max} > {30}$.}
Figure~\ref{fig:thetav_xf_map_outer} shows the results from such a calculation, displaying the photon number density flux map as functions of the scattered energy $x_f$ and the viewing colatitude $\theta_v$. Here {we have taken the same magnetic and observer inclination angles ($\alpha=0.25$ and $\beta = 1.0$) fitted from \emph{IXPE} data as in Section~\ref{sec:equatorial_spec}, and} the number density flux is integrated over the volume exterior to the field line with $\rmax ={30}$, corresponding to a footpoint angle $\theta_{\rm foot}<{0.18}$, and we adopt $r = 60$ as the outer integration boundary. This boundary is large enough that no photons of interest would resonantly scatter outside this surface. In this polar-dominated configuration, 
the scattered spectrum exhibits a lower cutoff energy when the line of sight approaches $\theta_v\approx\pi/2$, which is in contrast to the inner equatorial emission case shown in Figure~\ref{fig:thetav_xf_map}. The hardest power laws are achieved close to the two magnetic poles, but right at the pole resonance becomes difficult since photons collide with outflowing electrons tail-on, leading to a suppression around $\theta_v = 0$ and $\pi$.

We display the phase-resolved spectra for the $\rmax> 30$ configuration 
in Figure~\ref{fig:phase_resol_spectra_rm_gt_30}. %
As the rotational phase increases, the spectrum progressively softens and the hard X-ray component diminishes. 
This behavior is consistent with Figure~\ref{fig:thetav_xf_map_outer} where the softest spectra occur for viewing angle $\theta_v\sim\pi/2$. 
{We proceed to look at the phase-averaged spectra for a series of different polar twist configurations, where the boundary of the twist increases from $r_{\rm max} = 20$ all the way to $60$.} Figure~\ref{fig:phase_ave_spectra_rm_gt_20_30_60} {shows the computed phase-averaged spectra}. {We find that,} in general, adopting a larger $\rmax$ limit, or equivalently selecting a smaller polar footpoint angle $\theta_{\rm foot}$, yields a softer phase-averaged spectrum.

A pronounced thermal bump appears in all cases with $\perp$ mode seed photons. Again, this arises because we include a large equatorial region in the spectra integration, where the electrons are mostly non-relativistic and have higher density due to the {significant deceleration from resonant Compton drag},
and because we assume an unattenuated thermal seed-photon bath. In this regime, the cross section for the $\perp$ mode photon is essentially insensitive to the incoming angle between the photon momentum and
the magnetic field direction. By contrast, the $\parallel$ mode cross section carries a factor of $\cos^2{\theta_i}$, which reduces the cross section.

As shown in Figure~\ref{fig:r_reso_tau}, the magnetosphere can be optically thick to $\perp$ mode surface seed photons. Consequently, %
the scattered photon flux is no longer proportional to the electron number when the density is sufficiently high. In this case,
a self-consistent radiative-transfer calculation, including attenuation and possibly multiple scattering, is therefore required to model the soft X-ray emission reliably. %

{We note that the discrepancy of our results with the \emph{NuSTAR} data in Figures~\ref{fig:phase_resol_spectra_rm_gt_30} and~\ref{fig:phase_ave_spectra_rm_gt_20_30_60} can mostly be attributed to fixing the magnetic and viewing geometries. 
We have separately experimented with alternative dipole inclination angles and lines of sight that can reproduce a hard power law comparable to the \emph{NuSTAR} observations. 
This mismatch indicates that the polar-twist configuration adopted here is incompatible with the RVM constraints inferred from the \emph{IXPE} polarization data. Alternatively, it may suggest that a more realistic magnetic field configuration is required. Since the RCS calculation is very sensitive to the field configuration, a localized or asymmetric field configuration is expected to modify the resulting hard X-ray spectra. A joint fit to the \emph{IXPE}
polarization data, the \emph{NuSTAR} spectra, and the soft X-ray pulse profiles using a more flexible field configuration 
would therefore provide a powerful diagnostic of the magnetospheric structure. Such an analysis would allow the viewing geometry, particle distribution, and scattering
region to be constrained consistently, and represents a natural continuation of the current work.
\subsection{Pulsed Fraction, Pulse Profile, and Polarization}

Pulsed fraction, pulse profiles, and polarization information could provide complementary diagnostics for constraining the system geometry. {The pulsed fraction is defined as:}
\begin{equation}
    PF = \frac{F_{\rm max}-F_{\rm min}}{{F_{\rm max}+F_{\rm min}}},
\end{equation}
where $F_{\rm max}$ and $F_{\rm min}$ are the maximum and minimum flux in a given energy bin over different rotational phases. 

\begin{figure}
    \centering
    \includegraphics[width=0.95\linewidth]{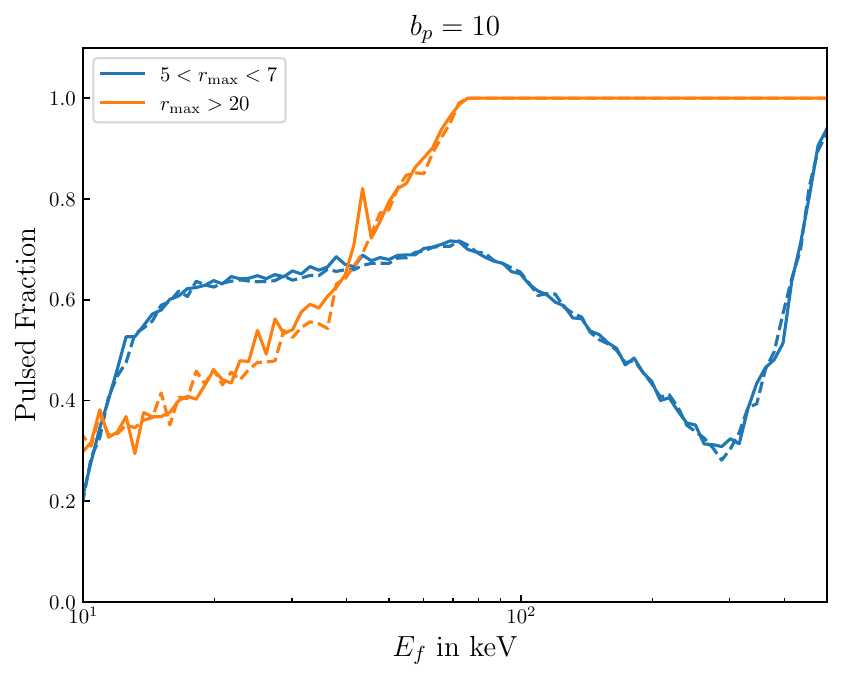}
    \caption{
    X-ray pulsed fraction PF plotted as a function of outgoing energy $E_f$. The blue curves display the PF for the surface seed photon case with $\perp$ (solid) or $\parallel$ (dashed) mode polarization. The orange curves depict the case with equatorial seed photons.}
    \label{fig:PF}
\end{figure}
{Figure~\ref{fig:PF} shows} the {energy dependent} pulsed fraction {of the two scenarios analyzed earlier}.
For the equatorial region with $5<\rmax<7$,
the calculated pulsed fraction rises with energy above around 10 keV, then declines at around 80 keV. The turnover is driven by a shift in the dominant spectral component across different rotational phases (see the blue and orange curves in Figure \ref{fig:phase_resol_spectra}). The pulsed fraction can also increase abruptly when the spectral cut-off is reached for one (or a subset) of the rotational phases. In the polar emission case with $\rmax>20$, the pulsed fraction increases monotonically with energy and approaches unity at around 80 keV, reflecting the disappearance of the hard X-ray component over part of the rotation cycle.
The predicted trend of the pulsed fraction is broadly consistent with observations, which show that the pulsed fraction generally increases with energy and can approach nearly 100\% in the hard X-ray band for several sources \citep[][]{2006ApJ...645..556K,2008A&A...489..263D,2008A&A...489..245D}. Nevertheless, some sources exhibit more complicated, non-monotonic or flatter behaviors \citep[e.g.,][]{2013ApJ...779..163A,2015ApJ...807...93A}, and quantitative comparisons depend on the specific definition of pulsed fraction adopted.

\begin{figure}
    \centering
    \includegraphics[width=0.95\linewidth]{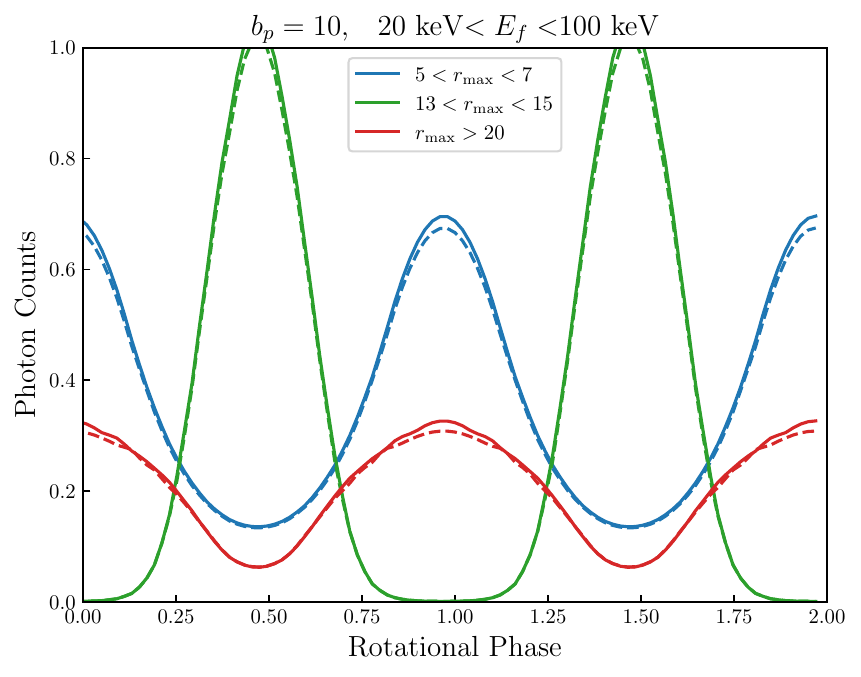}
    \caption{Pulse profiles of photon number counts integrated from 20 keV to 100 keV. The upper panel displays the pulse profiles for the $5<\rmax<7$ (blue), $13<\rmax<15$ (green), and $20<\rmax$ (red) cases, for both $\perp$ (solid) and $\parallel$ (dashed) mode seed photons, and the lower panel gives the polarization fraction for the same emission regions.}
    \label{fig:pulse_profiles}
\end{figure}

We now look at how the location of the twisted flux bundle affects the hard X-ray pulse profile. Figure~\ref{fig:pulse_profiles} gives the pulse profiles for scattered photons with final energy $E_f$ in the 20--100 keV band, for three different twist configurations: $5<\rmax<7$, $13<\rmax<15$, and $\rmax>20$.
{In computing these pulse profiles, we adopted the same magnetic and viewing angles inferred from \emph{IXPE} observations of source 4U 0142+61, $\alpha = 0.25$ and $\beta = 1.0$.}
As the emitting flux tube is extended outward, the pulse maximum systematically shifts in rotational phase $\phi$, evolving from 0 to $\pi$.
For emission regions near the polar cap with large $\rmax$, the peak subsequently shifts back toward $\phi\sim 0$.  %

The evolution of the pulse peaks reflects the interplay between the resonance condition and the viewing geometry. For an equatorial flux tube with small $\rmax$, the hard X-ray emission is dominated by scatterings near the top of the flux tube. In this case the maximum flux is achieved when the line of sight is close to the magnetic axis, and the strongest flux comes from the lower hemisphere, relativistically beamed in the upward direction. As $\rmax$ increases, the resonant energy near the flux tube apex decreases, so the hard X-ray photons originate preferentially at lower altitude, and the maximum flux occurs at phases where the line of sight intersects the brightest part of the tube at a finite angle to the magnetic axis. Finally, in the polar-region emission case, the scattered radiation is again beamed primarily along the magnetic-axis direction, which naturally returns the pulse maximum to $\phi\sim 0$. %
These distinct features of the pulse profiles can potentially be used to distinguish between different emission regions, even when all the different emission regions reproduce the observed hard X-ray power-law spectra (see Figure \ref{fig:phase_ave_spectra}). A systematic scan of the parameter space and fitting to the observed energy-dependent pulse profile is beyond the scope of this exploratory paper, and will be performed in a future work.
\begin{figure}
    \centering
    \includegraphics[width=0.95\linewidth]{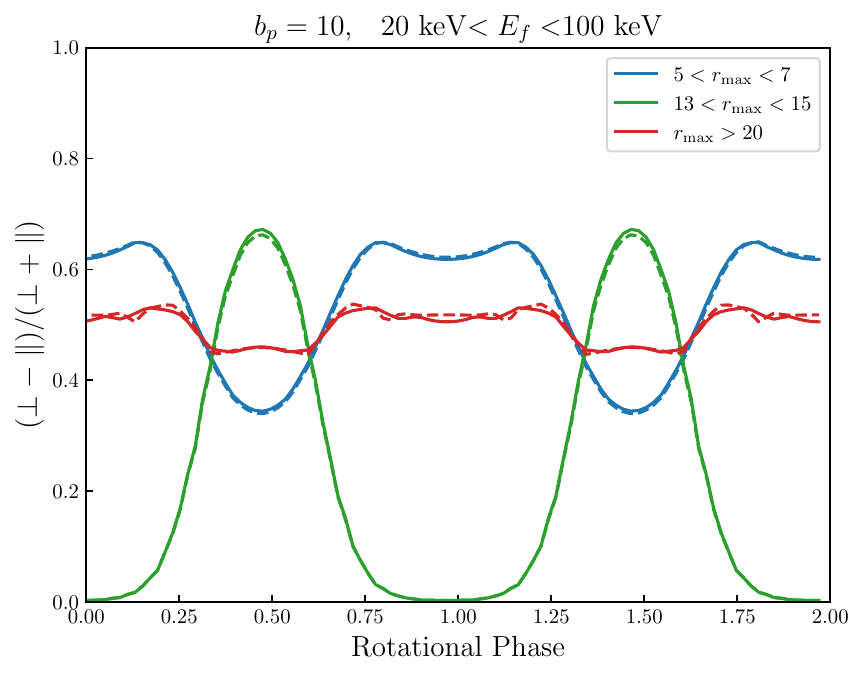}
    \includegraphics[width=0.95\linewidth]{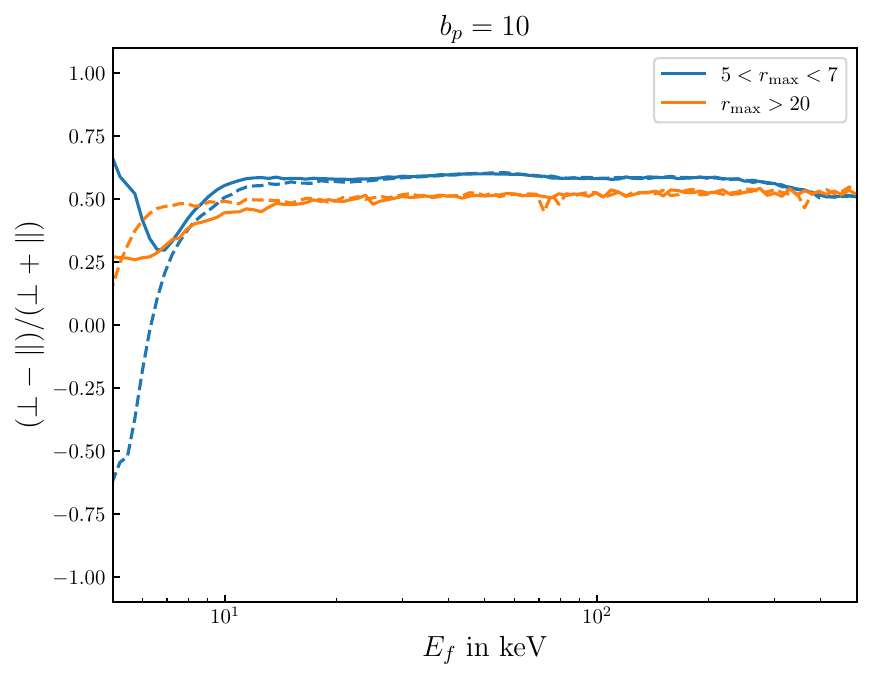}
    \caption{Upper panel: signed polarization fraction plotted as a function of the rotational phase. The emission regions are the same as those in Figure~\ref{fig:pulse_profiles}. Again, the solid and dashed curves represent the cases with $\perp$ and $\parallel$ mode seed photons, respectively. Lower panel:
    Signed polarization fraction plotted as a function of final energy $E_f$. }
    \label{fig:PD}
\end{figure}

The upper panel of Figure~\ref{fig:PD} shows the polarization fraction as a function of the rotational phase. 
In general, the scattered photons are predominantly polarized in the $\perp$ mode, and can reach a polarization fraction above 50\%. The polarization fraction typically peaks near the pulse maxima, i.e., it is broadly in phase with the pulse peaks. 
We also display the phase-averaged polarization degree as a function of the final energy $E_f$ in Figure~\ref{fig:PD}. Above around 10 keV, the polarization remains stable with a polarization degree of around 50\% to 60\%. This is consistent with theoretical expectations for a single RCS event, which yields a polarization fraction of roughly 0.5. Specifically, the integrated scattering cross sections realize a 3:1 ratio, where 3/4 of the scattered photons are in the $\perp$ mode and 1/4 of them in the $\parallel$ mode \citep[][]{2013ApJ...762...13B}. At low energies, both the flux and the polarization are dominated by surface photons. Therefore, in cases where the surface photons are initially in the $\parallel$ mode, a distinct polarization swing occurs once the $\perp$ mode RCS photons begin to dominate the observed flux. In the polar-twist geometry, this transition of the dominant polarization mode occurs at even lower energy, because scattering at higher altitudes corresponds to lower resonance energies. At hard X-ray energies, however, the polarization behavior becomes insensitive to the polarization mode of the seed photons. This result is also anticipated, because for highly relativistic electrons, incident photons in the $\perp$ and $\parallel$ modes become effectively degenerate in the ERF.
%


\section{Conclusions and Discussion}
\label{sec:conclusion}

In this paper, we systematically explored the consequences of resonant Compton scattering (RCS) of X-ray photons in a twisted magnetar magnetosphere. We first computed how electrons decelerate due to RCS, and inferred the magnetospheric opacity for thermal X-ray photons emitted from the star. Then, we specialized to two twist configurations and computed the resulting hard X-ray emission properties.
\red{The primary advancement of our model is the consistent combination of the pair-outflow geometry, the RCS pair deceleration, and the hard X-ray production, alongside the coupling of the hard X-ray flux to the soft X-ray opacity of the magnetosphere.}
We summarize our main conclusions as follows:

\begin{enumerate}
    \item Assuming an acceleration and pair production region very close to the stellar surface, we self-consistently computed the deceleration of electrons and positrons caused by RCS in the twisted magnetosphere. We found that RCS cooling significantly increases the electron density and renders a large fraction of the magnetosphere optically thick to $\perp$ mode thermal X-ray photons from the surface.
    \item To circumvent the optical depth issue, we propose three scenarios: a twisted flux tube located in the equatorial region very close to the star; a twisted polar region; or thermal photons from the star that are primarily emitted in the $\parallel$ mode \footnote{\red{For possible explanations of this polarized surface emission behavior, see \citet{2023PNAS..12016534L} for mode conversion at the vacuum resonance in a gaseous atmosphere, or  \cite{2020MNRAS.492.5057T} for the emission from a condensed surface.}}.
    \item We computed the RCS hard X-ray spectra for the two types of twist configurations that can avoid an optically thick magnetosphere. In both cases, we adopted the magnetic inclination and viewing angles inferred from comparing the \emph{IXPE} observations of 4U 0142+61 with the rotating vector model. We found that the polar-twist configuration cannot reproduce the hard X-ray power law observed by \emph{NuSTAR}, while the equatorial-twist configuration can produce spectra broadly consistent with the observations.
    \item We also computed the hard X-ray pulsed fraction, pulse profiles, and polarization properties. 
    We find that the peaks of the hard X-ray pulse profile can have a shift with respect to the soft X-ray peaks depending on the location of the twisted field line bundle. 
    The hard X-ray pulsed fraction generally increases with energy. Furthermore, scattering can produce a hard X-ray polarization fraction of approximately 50\%, providing a key diagnostic to verify the RCS scenario.
\end{enumerate}

In this paper, we adopted the classical magnetic Thomson scattering cross section and approximated the cyclotron resonance with a delta function. Although this delta function treatment can overestimate the cross section in the strong field regime, it does not change our qualitative conclusions. We verified our results by repeating the calculation using a more accurate recipe for the cyclotron line width in Equation~(\ref{eqn:x_Gamma}), and the resulting spectra below 1 MeV differ only slightly (typically by less than a few percent). This is because the quantum correction to the line width starts to be important when $b$ approaches or exceeds unity. However, photons with $E_f$ around a few hundred keV are primarily scattered in sub-critical field regions, and the cooling of the electrons also becomes prominent in the sub-critical regime. Therefore, the classical approximation of the cross section is adequate for our purposes. Inclusion of a more accurate cross section and full relativistic kinematics is deferred to future work. 

We did not include photon splitting in the present calculations. This omission is not expected to affect our main results, since photon splitting is not significant for the $B_p = 10$ case unless photons are injected toward the star \citep[see Figures 5 and 9 of ][]{2019MNRAS.486.3327H}. In this work, we only consider the outflow mode where scattered photons are beamed outward. Under such conditions, photon splitting would mainly influence the scattered high-energy photons in the very inner magnetosphere and affect pair production rates, but has little impact on the hard X-ray photons generated by the outflowing plasma.

Although the above assumptions limit the quantitative interpretation of some details, they do not alter the main findings of this work. 
\red{Our calculation assumes the seed surface emission is fully polarized in either $\perp$ or $\parallel$ mode. However, in the ultra-strong magnetic fields of magnetars, QED effects introduce an additional layer of complexity. 
As soft X-ray photons propagate outward in the surface layer, they cross the vacuum resonance layer where adiabatic mode conversion can occur \citep[e.g., see][]{2003ApJ...588..962L,2023PNAS..12016534L,2026arXiv260308119G}. This process can significantly alter the polarization of the surface emission, potentially yielding a $\perp$-polarized spectrum at lower energies and a $\parallel$-polarized one at slightly higher energies. 
The recently observed polarization degree dip of the magnetar 1E 1547.0-5408 indicates the performance of partial mode conversion at the vacuum resonance \citep[][]{2026ApJ..1002..102T}. Crucially, because the electron deceleration dynamics and the resulting inverse Compton spectrum are largely insensitive to the initial polarization state of the seed photons, this QED vacuum resonance does not impact our core hard X-ray calculations.}

\red{
It does, however, profoundly impact the emergent soft X-ray polarization. This interplay between QED mode conversion and magnetospheric scattering further complicates the interpretation of phase-resolved polarization measurements by \emph{IXPE}. }
In particular, for a large portion of the magnetosphere, the calculated radial optical depth is high for the surface thermal soft X-ray emission in the $\perp$ mode. Depending on the shape of the emission region, the optical depth could far exceed unity, especially for photons propagating in the equatorial directions. Therefore, %
phase-resolved surface flux and polarization measurements could strongly constrain the viewing and emission geometries.
On the other hand, 
if the optical depth is indeed large, the surface soft X-ray photons would be reprocessed before escaping the magnetosphere. Even in this high optical-depth regime, most soft photons are expected to scatter only a few times, because each scattering modifies the photon energy and momentum and therefore changes whether the resonance condition can still be satisfied. As a result, the reprocessed photons should largely retain a quasi-thermal spectral shape, but have distinct polarization signatures.
This might have a profound impact on interpreting the soft X-ray polarization signals observed by \emph{IXPE}. 
\red{Extensive efforts have been made on the modeling of repeated RCS spectra in the moderate opacity regime through Monte Carlo radiative transfer techniques \citep{2007ApJ...660..615F,2008MNRAS.386.1527N,2011ApJ...730..131F,2014MNRAS.438.1686T}. Additionally, semi-analytical treatments in the single scattering regime have shown that mode-switch features in the \emph{IXPE} band arise due to both vacuum resonance and RCS \citep[][]{2026arXiv260308119G}. 
Building upon these foundational models, understanding}
the opacity and radiative transfer of the surface thermal photons \red{in the optically thick regime} is essential to interpret the polarization data from \emph{IXPE} and will be addressed in future work.
Some of our spectra display an enhanced quasi-thermal soft X-ray component for $\perp$ mode seed photons. %
This enhancement arises because the RCS spectra are calculated in the linear, optically thin regime (see Equation~\ref{eqn:dNph_dtdxdOdNe},  and also \citealp{2007Ap&SS.308..109B, 2018ApJ...854...98W}). In this limit, the outgoing flux is proportional to the electron density at the scattering site. This approximation is valid when the electron density is sufficiently low that all electrons have a comparable probability of being illuminated by the incident photons at the scattering site.
However, in the equatorial region of models with large $\rmax$, decelerated electrons can accumulate and make the magnetosphere optically thick to surface $\perp$ mode photons. %
In this regime, the outgoing flux no longer scales linearly with the electron density, and the resulting soft X-ray enhancement is therefore not physical.
For $\parallel$ mode seed photons, the scattering cross section is suppressed by a factor of $\cos^2{\theta'_i}$. Near the equatorial region, the incident angle approaches $\pi/2$ for non-relativistic electrons, so the scattering probability is significantly reduced. 
The soft X-ray photons emerging from the optically thick region may also scatter back into the optically thin region and undergo additional RCS with the relativistic electrons, further complicating the electron cooling and the radiative transfer problem.
A detailed radiative transfer calculation, for example using Monte Carlo techniques or the Sobolev method, together with self-consistent electron cooling, is required to model the soft X-ray emission. This is likely the crucial next step for the proper theoretical interpretation of the steep power law observed in the 
soft X-ray range.

The hard X-ray emission possesses strong polarization, which generally exceeds 50\% above 10 keV. This makes them a perfect target for hard X-ray instruments like {\it XL-Calibur} or the \emph{Compton Spectrometer and Imager} (\emph{COSI}). A phase-averaged polarization measurement could confirm the RCS mechanism and help to pinpoint the polarization mode for the soft X-ray emission, while phase-resolved measurements are required to probe the scattering geometry.

\begin{acknowledgments}
\red{We thank the anonymous referee for valuable comments and constructive suggestions that improved the paper.}
We thank Matthew Baring, William Charles, Alice Harding, Henric Krawczynski, John Mehlhaff, Mike Nowak, Zorawar Wadiasingh, and Yajie Yuan for helpful discussions. This work is supported by NASA grant 80NSSC24K1095.  A.C. additionally acknowledges support from NSF grants DMS-2235457 and AST-2308111, as well as NASA grant 80NSSC25K0080. The authors also acknowledge the Research Infrastructure Services (RIS) group at Washington University in St. Louis for providing computational resources and services that were used to generate parts of the research results delivered within this paper.
\end{acknowledgments}

\appendix
{\color{black}

\section{The Scattering Transfer}
\label{app:scattering}
Here we briefly discuss the scattering transfer equation and its solution in the optically thin regime. we formulate the radiative transfer equation in terms of the specific photon number intensity, $\mathcal{N}=dn/(dx d\Omega)$, which can be connected to the observed number flux at energy $xm_ec^2$ via 
\begin{equation}
    F_N(x) = \frac{1}{D^2} \int_{A} \mathcal{N}(x, \mathbf{\Omega}_{\rm obs}) \, dA,
\end{equation}
where $D$ is the distance of the source star and $A$ is the projected area of the source perpendicular to the line of sight. 
The relativistic scattering transfer equation along a path length $s$ in steady state is given by:
\begin{equation}
\frac{d\mathcal{N}_{q}(\mathbf{\Omega}_f)}{ds} = - n_e (1 - \beta \mu_f) \sigma'_q(x'_f) \mathcal{N}_{q}(\mathbf{\Omega}_f) + \frac{n_e}{\gamma(1 - \beta \mu_f)} \sum_p\int d\Omega_i \int dx_i (1 - \beta \mu_i) \frac{d\sigma'(x'_i \to x'_f, \mathbf{\Omega}'_i \to \mathbf{\Omega}'_f)}{dx'_f \, d\Omega'_f} \mathcal{N}_{p}(\mathbf{\Omega}_i),
\label{eqn:relativistic_RT}
\end{equation}
where $n_e$ is the local $e^\pm$ number density, and $\mu_i$ and $\mu_f$ are the cosines of the angles between the respective photon directions and the electron velocity vector in the laboratory frame. The subscripts $p$ and $q$ denote the polarization states of the incident and outgoing photons. Quantities evaluated in the ERF are denoted with primes. Equation~(\ref{eqn:relativistic_RT}) is rigorous unless the number densities of the electrons or photons are so extreme that the plasma frequency approaches the photon frequency, or the stimulated scattering becomes important. 

Equation~(\ref{eqn:relativistic_RT}) is a standard first-order linear differential equation. The general solution can be expressed in terms of the optical depth $\tau_q=\int n_e(1-\beta\mu_f)\sigma'_q ds$, yielding:
\begin{equation}
\mathcal{N}_{q}(s) = \mathcal{N}_q(s_0) e^{-\tau_q(s_0, s)} + \int_{s_0}^{s} j_q(s') e^{-\tau_q(s', s)} ds',
\label{eqn:general_solution}
\end{equation}
where
\begin{equation}
j_q(s) = \frac{n_e}{\gamma(1 - \beta \mu_f)} \sum_{p} \int d\Omega_i \int dx_i (1 - \beta \mu_i) \frac{d\sigma'_{p \to q}}{dx'_f \, d\Omega'_f} \mathcal{N}_p(s, \mathbf{\Omega}_i)
\label{eqn:effective_emissivity}
\end{equation}
is the effective emissivity. 
Here Equation~(\ref{eqn:effective_emissivity}) is identical to Equation~(\ref{eqn:dNph_dtdxdOdNe}) in the main text with an extra factor of the speed of light $c$. 
The general solution in Equation~(\ref{eqn:general_solution}) integrates the scattered-in photons emissivity ($j^q$) attenuated by the optical depth between the scattering point $s'$ and the observer at $s$, plus the initial boundary condition at $s_0$ (the neutron star surface) attenuated by the total optical depth $\tau_q(s_0, s)$.

In the optically thin regime, the attenuation of the intrinsic surface emission vanishes, and the incident photons in the integrand of the emissivity can be approximated by the unscattered seed photons coming straight from the surface, which we denote as $\mathcal{N}_{{\rm surf},p}$. Applying these limits, the general solution simplifies into the sum of the bare surface emission and an integral of the single scattered photons:
\begin{equation}
    \mathcal{N}_{q}(s) \approx \mathcal{N}_{{\rm surf},q}(x_f, \mathbf{\Omega}_f) + \int_{s_0}^{s} j_{{\rm thin},q}(s') ds'
    \label{eqn:thin_solution},
\end{equation}
where $j_{{\rm thin},q}$ is evaluated using the surface seed photons,
\begin{equation}
    j_{{\rm thin},q}(s') = \frac{n_e}{\gamma(1 - \beta \mu_f)} \sum_{p} \int d\Omega_i \int dx_i (1 - \beta \mu_i) \frac{d\sigma'_{p \to q}}{dx'_f \, d\Omega'_f} \mathcal{N}_{{\rm surf},p}(x_i, \mathbf{\Omega}_i).
    \label{eqn:thin_emissivity}
\end{equation}
The specific number intensity and the effective emissivity in the optically thin regime form the foundation of the calculation in the main text. Equation~(\ref{eqn:thin_solution}) recovers the unscattered surface emission in the limit of $n_e\rightarrow 0$. For hard X-ray emission ($x_fm_ec^2>10$ keV), the first term on the right-hand side of Equation~(\ref{eqn:thin_solution}) vanishes due to the low surface temperature, and the observed flux becomes a straightforward volume integration over the emissivity, resulting in Equation~(\ref{eqn:spec_vol_inte}) in the main text. 

\section{Reliability of the optical thin single scattering approximation}
\label{app:optical_thin}
Our calculation of the hard X-ray spectra relies on Equations~(\ref{eqn:thin_solution}) and (\ref{eqn:thin_emissivity}), which are derived from the general solution under the optically thin condition. 
However, the ``optically thin" assumption actually entails three distinct simplifications to Equations~(\ref{eqn:general_solution}) and (\ref{eqn:effective_emissivity}):
\begin{enumerate}

\item The attenuation of the source emission vanishes, therefore $e^{-\tau_q(s_0, s)}\approx1$.

\item The attenuation (or amplification) of the incident radiation from other directions vanishes, therefore the specific number intensity in the scattering kernel can be approximated by the source emission from the stellar surface, i.e., $\mathcal{N}_p\approx \mathcal{N}_{{\rm surf},p}$.

\item the scatter-in emission does not undergo further attenuation from the scattering site to the observer, therefore $e^{-\tau_q(s', s)}\approx 1$.
\end{enumerate}

In the context of RCS in a magnetar's magnetosphere, these three conditions may not all hold simultaneously at all energies. For example, if a radiation beam emitted from the star toward the observer does not intersect with any scattering region, conditions 1 and 3 are automatically satisfied. However, condition 2 could still break if a separate part of the magnetosphere is optically thick and reflects photons back into the path of the beam.

This paper focuses primarily on the RCS hard X-ray spectra. Condition 1 is trivially satisfied because the flux of the quasi-thermal surface emission vanishes in the hard X-ray band. Furthermore, the scattered hard X-rays are strongly beamed along the local magnetic field direction and propagate approximately outward, experiencing a rapidly decreasing magnetic field strength. In this geometry, it is unlikely for the hard X-ray photons to reach the resonance again, making condition 3 a reliable approximation. Condition 2 can also be satisfied provided the scattering region is confined to the polar region or the inner equatorial region (see the optical depth plots in Figure~\ref{fig:r_reso_tau}). Outside these zones, the deceleration of the $e^\pm$ pairs can make part of the magnetosphere optically thick to soft X-ray photons, and the reflected photons would subsequently contribute to the incident $\mathcal{N}_p$ in the scattering kernel. Therefore, our optically thin approximation is valid for the hard X-ray calculations as long as the active scattering region does not extend into the outer equatorial region ($\rmax\gtrsim 10$ for $B_p=10$).

For soft X-ray photons, conditions 2 and 3 can remain satisfied under the same geometric constraint (avoiding the outer equatorial region). Condition 1 is satisfied provided the direct radiation beam from the surface to the observer avoids intersecting the active scattering region.
}

\bibliography{sample7}{}
\bibliographystyle{aasjournalv7}



\end{document}